\documentclass{PoS}

\usepackage{cite} % Allows for ranges in citations
\usepackage{mciteplus}

\usepackage[percent]{overpic}

\usepackage{mathptmx}
\usepackage[scaled=.99]{helvet}
\usepackage{courier}
\DeclareMathAlphabet{\mathcal}{OMS}{cmsy}{m}{n}

\usepackage{lineno}
\usepackage{ifthen} % for conditional statements
\newboolean{pdflatex}
\setboolean{pdflatex}{true} % False for eps figures 

\newboolean{articletitles}
\setboolean{articletitles}{true} % False removes titles in references

\newboolean{uprightparticles}
\setboolean{uprightparticles}{false} %True for upright particle symbols

\newboolean{inbibliography}
\setboolean{inbibliography}{false} %True once you enter the
\usepackage{xspace} 
\usepackage{upgreek}

\def\MagUp {\mbox{\em Mag\kern -0.05em Up}\xspace}

\ifthenelse{\boolean{uprightparticles}}%
{

 \def\Pmu         {\ensuremath{\upmu}\xspace}                 
 \def\Pnu         {\ensuremath{\upnu}\xspace}                 
                  
 \def\Ppi         {\ensuremath{\uppi}\xspace}

 \def\Ppsi        {\ensuremath{\uppsi}\xspace}

 \def\PDelta      {\ensuremath{\Delta}\xspace}                 
 \def\PXi         {\ensuremath{\Xi}\xspace}                 
 \def\PLambda     {\ensuremath{\Lambda}\xspace}                 
 \def\PSigma      {\ensuremath{\Sigma}\xspace}                 
 \def\POmega      {\ensuremath{\Omega}\xspace}                 
 \def\PUpsilon    {\ensuremath{\Upsilon}\xspace}

 \def\PB      {\ensuremath{\mathrm{B}}\xspace}                 
                  
 \def\PD      {\ensuremath{\mathrm{D}}\xspace}

 \def\PJ      {\ensuremath{\mathrm{J}}\xspace}                 
 \def\PK      {\ensuremath{\mathrm{K}}\xspace}

 \def\Pb      {\ensuremath{\mathrm{b}}\xspace}                 
 \def\Pc      {\ensuremath{\mathrm{c}}\xspace}

 \def\Pi      {\ensuremath{\mathrm{i}}\xspace}

 \def\Pp      {\ensuremath{\mathrm{p}}\xspace}

 \def\Ps      {\ensuremath{\mathrm{s}}\xspace}

 \def\thebaroffset{0.0em}
}
{

 \def\Pmu         {\ensuremath{\mu}\xspace}                 
 \def\Pnu         {\ensuremath{\nu}\xspace}                 
                  
 \def\Ppi         {\ensuremath{\pi}\xspace}

 \def\Ppsi        {\ensuremath{\psi}\xspace}                 
                  
 \mathchardef\PDelta="7101
 \mathchardef\PXi="7104
 \mathchardef\PLambda="7103
 \mathchardef\PSigma="7106
 \mathchardef\POmega="710A
 \mathchardef\PUpsilon="7107
                  
 \def\PB      {\ensuremath{B}\xspace}                 
                  
 \def\PD      {\ensuremath{D}\xspace}

 \def\PJ      {\ensuremath{J}\xspace}                 
 \def\PK      {\ensuremath{K}\xspace}

 \def\Pb      {\ensuremath{b}\xspace}                 
 \def\Pc      {\ensuremath{c}\xspace}

 \def\Pi      {\ensuremath{i}\xspace}

 \def\Pp      {\ensuremath{p}\xspace}

 \def\Ps      {\ensuremath{s}\xspace}

 \def\thebaroffset{0.18em}
}
\newcommand{\offsetoverline}[2][\thebaroffset]{\kern #1\overline{\kern -#1 #2}}%

\makeatletter
\ifcase \@ptsize \relax% 10pt
  \newcommand{\miniscule}{\@setfontsize\miniscule{4}{5}}% \tiny: 5/6
\or% 11pt
  \newcommand{\miniscule}{\@setfontsize\miniscule{5}{6}}% \tiny: 6/7
\or% 12pt
  \newcommand{\miniscule}{\@setfontsize\miniscule{5}{6}}% \tiny: 6/7
\fi
\makeatother

\DeclareRobustCommand{\optbar}[1]{\shortstack{{\miniscule (\rule[.5ex]{1.25em}{.18mm})}
  \\ [-.7ex] $#1$}}

\def\mun        {{\ensuremath{\Pmu^-}}\xspace} % muon negative (\mum is taken)

\def\neub       {{\ensuremath{\overline{\Pnu}}}\xspace}

\def\neumb      {{\ensuremath{\neub_\mu}}\xspace}
\def\squark    {{\ensuremath{\Ps}}\xspace}

\def\cquark    {{\ensuremath{\Pc}}\xspace}

\def\bquark    {{\ensuremath{\Pb}}\xspace}

\def\pion   {{\ensuremath{\Ppi}}\xspace}

\def\pip    {{\ensuremath{\pion^+}}\xspace}
\def\pim    {{\ensuremath{\pion^-}}\xspace}

\def\kaon    {{\ensuremath{\PK}}\xspace}
\def\KorKbar {\kern \thebaroffset\optbar{\kern -\thebaroffset \PK}{}\xspace}

\def\Kp      {{\ensuremath{\kaon^+}}\xspace}
\def\Km      {{\ensuremath{\kaon^-}}\xspace}

\def\D       {{\ensuremath{\PD}}\xspace}

\def\DorDbar {\kern \thebaroffset\optbar{\kern -\thebaroffset \PD}\xspace}
\def\Dz      {{\ensuremath{\D^0}}\xspace}

\def\Dp      {{\ensuremath{\D^+}}\xspace}
\def\Dm      {{\ensuremath{\D^-}}\xspace}

\def\DpDm    {\ensuremath{\Dp {\kern -0.16em \Dm}}\xspace}

\def\Ds      {{\ensuremath{\D^+_\squark}}\xspace}
\def\Dsp     {{\ensuremath{\D^+_\squark}}\xspace}

\def\B       {{\ensuremath{\PB}}\xspace}

\def\BorBbar {\kern \thebaroffset\optbar{\kern -\thebaroffset \PB}\xspace}

\def\Bd      {{\ensuremath{\B^0}}\xspace}

\def\BdorBdbar {\kern \thebaroffset\optbar{\kern -\thebaroffset \Bd}\xspace}
\def\Bu      {{\ensuremath{\B^+}}\xspace}

\def\Bs      {{\ensuremath{\B^0_\squark}}\xspace}

\def\BsorBsbar {\kern \thebaroffset\optbar{\kern -\thebaroffset \Bs}\xspace}
\def\Bc      {{\ensuremath{\B_\cquark^+}}\xspace}

\def\jpsi     {{\ensuremath{{\PJ\mskip -3mu/\mskip -2mu\Ppsi}}}\xspace}

\def\Y#1S{\ensuremath{\PUpsilon{(#1S)}}\xspace}

\def\proton      {{\ensuremath{\Pp}}\xspace}
\def\antiproton  {{\ensuremath{\overline \proton}}\xspace}

\def\Lz          {{\ensuremath{\PLambda}}\xspace}
\def\Lbar        {{\ensuremath{\offsetoverline{\PLambda}}}\xspace}
\def\LorLbar     {\kern \thebaroffset\optbar{\kern -\thebaroffset \PLambda}\xspace}

\def\Xires       {{\ensuremath{\PXi}}\xspace}

\def\Omegares    {{\ensuremath{\POmega}}\xspace}

\def\Lc          {{\ensuremath{\Lz^+_\cquark}}\xspace}

\def\Xic         {{\ensuremath{\Xires_\cquark}}\xspace}
\def\Xicz        {{\ensuremath{\Xires^0_\cquark}}\xspace}
\def\Xicp        {{\ensuremath{\Xires^+_\cquark}}\xspace}

\def\Omegac      {{\ensuremath{\Omegares^0_\cquark}}\xspace}

\def\Xiccp       {{\ensuremath{\Xires^+_{\cquark\cquark}}}\xspace}
\def\Xiccpp      {{\ensuremath{\Xires^{++}_{\cquark\cquark}}}\xspace}

\def\Lb           {{\ensuremath{\Lz^0_\bquark}}\xspace}
\def\Lbbar        {{\ensuremath{\Lbar{}^0_\bquark}}\xspace}

\def\Xib          {{\ensuremath{\Xires_\bquark}}\xspace}
\def\Xibz         {{\ensuremath{\Xires^0_\bquark}}\xspace}
\def\Xibm         {{\ensuremath{\Xires^-_\bquark}}\xspace}

\def\Omegab       {{\ensuremath{\Omegares^-_\bquark}}\xspace}

\def\to                 {\ensuremath{\rightarrow}\xspace}

\def\AT#1     {\ensuremath{A_{\mathrm{T}}^{#1}}\xspace}           % 2

\def\C#1      {\ensuremath{\mathcal{C}_{#1}}\xspace}                       % 9
\def\Cp#1     {\ensuremath{\mathcal{C}_{#1}^{'}}\xspace}                    % 7
\def\Ceff#1   {\ensuremath{\mathcal{C}_{#1}^{\mathrm{(eff)}}}\xspace}        % 9  
\def\Cpeff#1  {\ensuremath{\mathcal{C}_{#1}^{'\mathrm{(eff)}}}\xspace}       % 7
\def\Ope#1    {\ensuremath{\mathcal{O}_{#1}}\xspace}                       % 2
\def\Opep#1   {\ensuremath{\mathcal{O}_{#1}^{'}}\xspace}                    % 7

\newcommand{\tev}{\ifthenelse{\boolean{inbibliography}}{\ensuremath{~T\kern -0.05em eV}}{\ensuremath{\mathrm{\,Te\kern -0.1em V}}}\xspace}
\newcommand{\gev}{\ensuremath{\mathrm{\,Ge\kern -0.1em V}}\xspace}
\newcommand{\mev}{\ensuremath{\mathrm{\,Me\kern -0.1em V}}\xspace}
\newcommand{\kev}{\ensuremath{\mathrm{\,ke\kern -0.1em V}}\xspace}
\newcommand{\ev}{\ensuremath{\mathrm{\,e\kern -0.1em V}}\xspace}
\newcommand{\mevc}{\ensuremath{{\mathrm{\,Me\kern -0.1em V\!/}c}}\xspace}
\newcommand{\gevc}{\ensuremath{{\mathrm{\,Ge\kern -0.1em V\!/}c}}\xspace}
\newcommand{\mevcc}{\ensuremath{{\mathrm{\,Me\kern -0.1em V\!/}c^2}}\xspace}
\newcommand{\gevcc}{\ensuremath{{\mathrm{\,Ge\kern -0.1em V\!/}c^2}}\xspace}
\newcommand{\gevgevcc}{\ensuremath{{\mathrm{\,Ge\kern -0.1em V^2\!/}c^2}}\xspace} % for \pt^2 in CEP
\newcommand{\gevgevcccc}{\ensuremath{{\mathrm{\,Ge\kern -0.1em V^2\!/}c^4}}\xspace} % for q^2

\def\gsim{{~\raise.15em\hbox{$>$}\kern-.85em
          \lower.35em\hbox{$\sim$}~}\xspace}
\def\lsim{{~\raise.15em\hbox{$<$}\kern-.85em
          \lower.35em\hbox{$\sim$}~}\xspace}

\def\sqs   {\ensuremath{\protect\sqrt{s}}\xspace}

\def\pt         {\ensuremath{p_{\mathrm{T}}}\xspace}

\def\pythia     {\mbox{\textsc{Pythia}}\xspace}

\def\tell1  {TELL1\xspace}
\def\ukl1   {UKL1\xspace}

\title{Heavy flavour production and spectroscopy}

\ShortTitle{Heavy flavour production and spectroscopy}

\author{\speaker{Jibo He}\thanks{on behalf of the ALICE, ATLAS, CMS
    and LHCb collaborations.}\\
        University of Chinese Academy of Sciences (UCAS), Beijing,
        China\\
        E-mail: \email{jibo.he@cern.ch}}

\abstract{Latest results on the heavy flavour production and
  spectroscopy at the LHC are reviewed. These include
  measurements of production rates of the charmed and beauty hadrons, and 
  observations of new excited charmed and beauty hadrons and exotic states.}

\FullConference{%
 
 The Ninth Annual Conference on Large Hadron Collider Physics - LHCP2021\\
 7-12 June 2021\\
 Online
}

\begin{document}

\vskip-0.8cm
\section{Introduction}
\vskip-0.2cm

Study of heavy flavour production and spectroscopy is important to
understand the strong interaction.
Large production rates of the heavy flavour particles at the LHC open
the doors to extend analyses to unexplored phase spaces, and bring new
information to help understand the QCD. More than fifty new conventional or exotic states 
have been observed at the LHC.  
Precise measurements of heavy
flavour production cross-sections and distributions are important to
test the theoretical predictions being used to estimate backgrounds to
new physics searches. Measurements of heavy flavour production
in proton-proton ($pp$) collisions
also provide a reference for those in heavy-ion collisions to understand
nuclear effects. 

\vskip-0.8cm
\section{Heavy flavour production}
\vskip-0.2cm

\begin{figure}[b]
  \vskip-0.4cm
  \centering
  \resizebox{0.32\textwidth}{!}{%
    \includegraphics{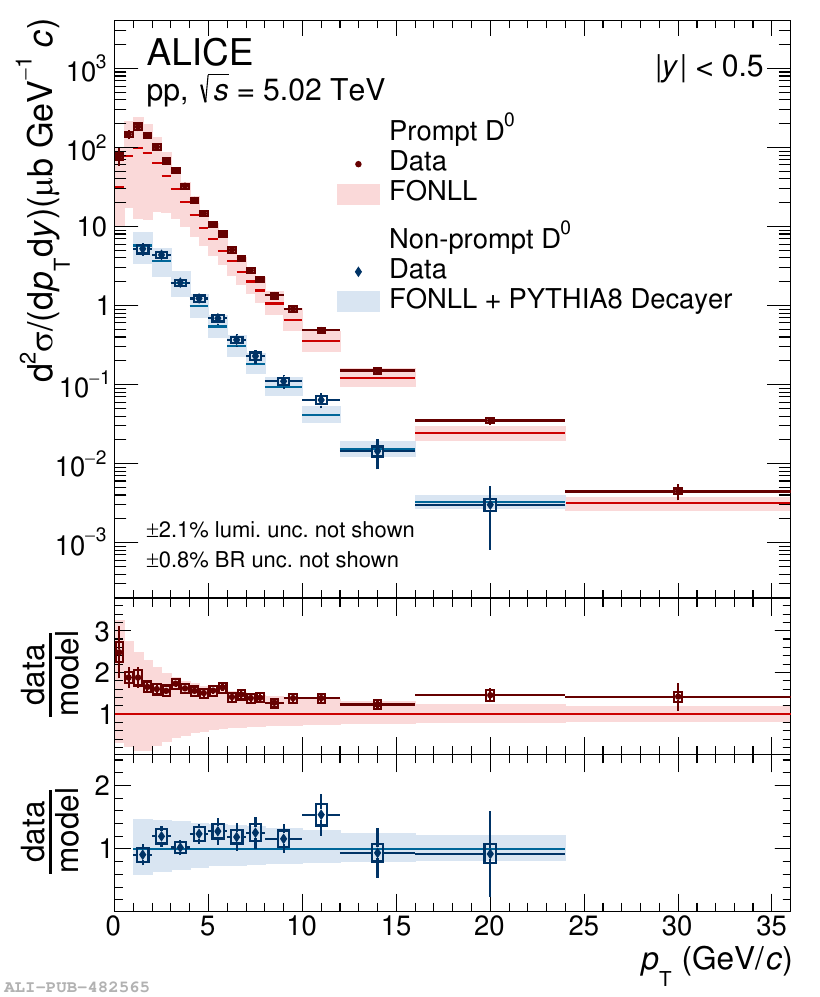}}
  \resizebox{0.32\textwidth}{!}{%
    \includegraphics{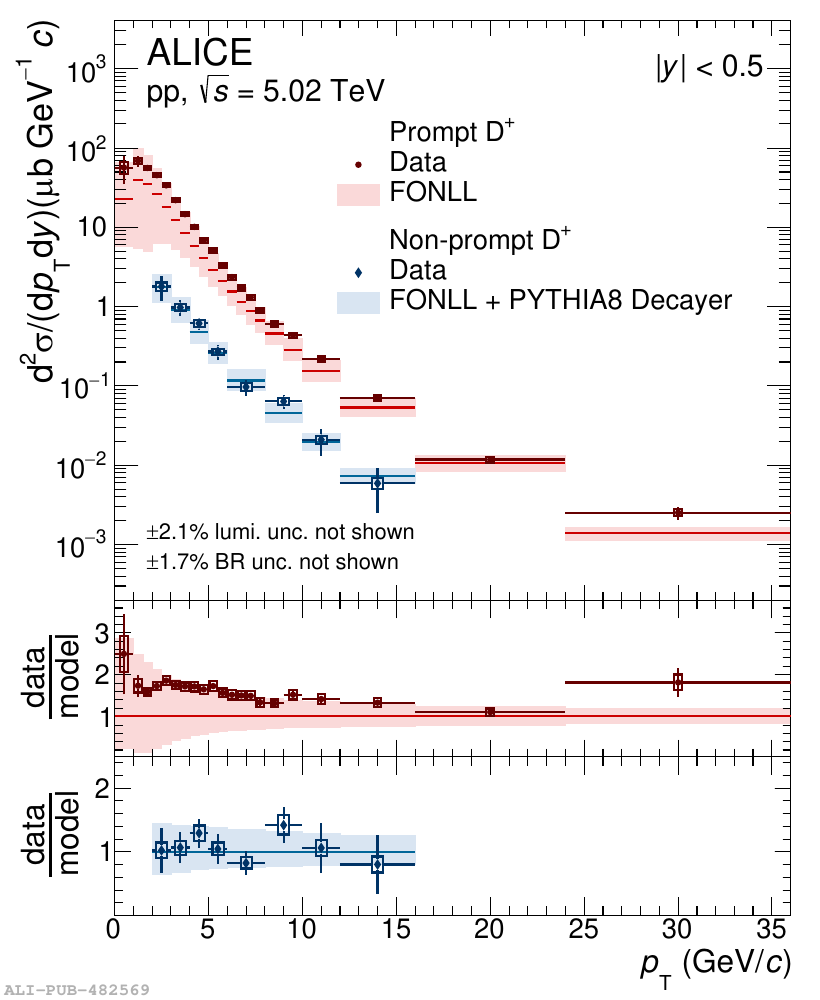}}
  \resizebox{0.32\textwidth}{!}{%
    \includegraphics{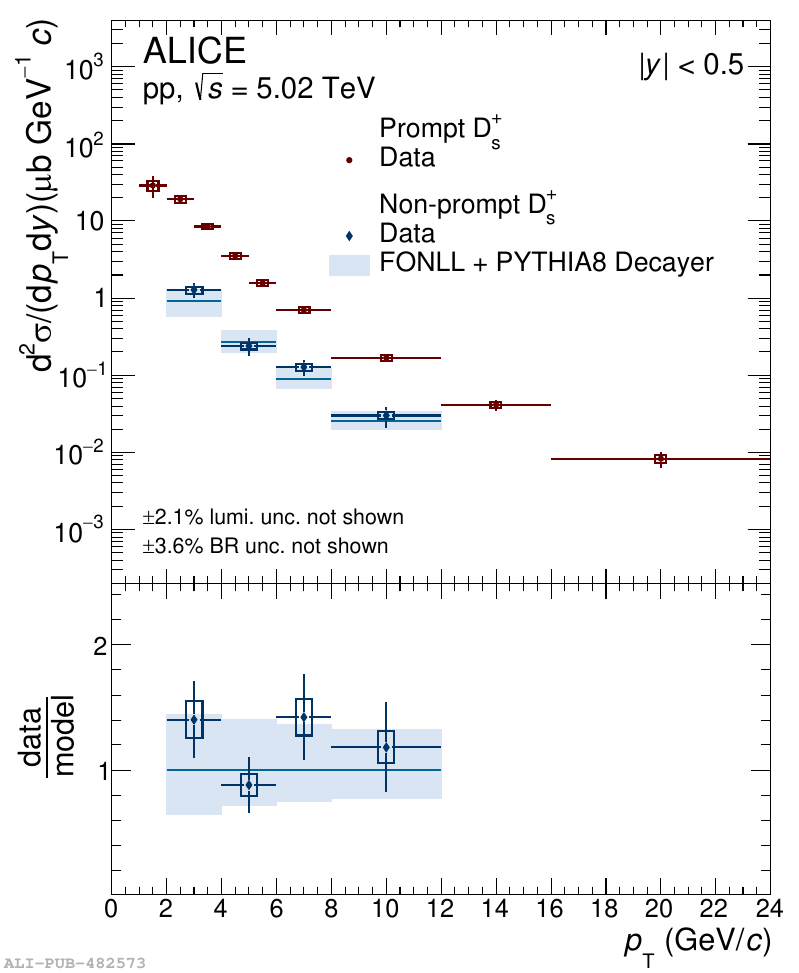}}
  \vskip-0.2cm
  \caption{Production cross-sections of prompt and non-prompt $\Dz,
    \Dp, \Ds$ mesons~\cite{ALICE:2021mgk} compared to predictions obtained with FONLL
    calculations~\cite{Cacciari:1998it,Cacciari:2001td},
    combined with \pythia8~\cite{Sjostrand:2006za,Sjostrand:2014zea} for
    the $H_b\to D+X$ decay kinematics.}
  \label{fig:AliceDVsFONLL}
  \vskip-0.4cm
\end{figure}

\begin{figure}
  \vskip-0.5cm
  \centering
  \resizebox{0.45\textwidth}{!}{%
    \includegraphics{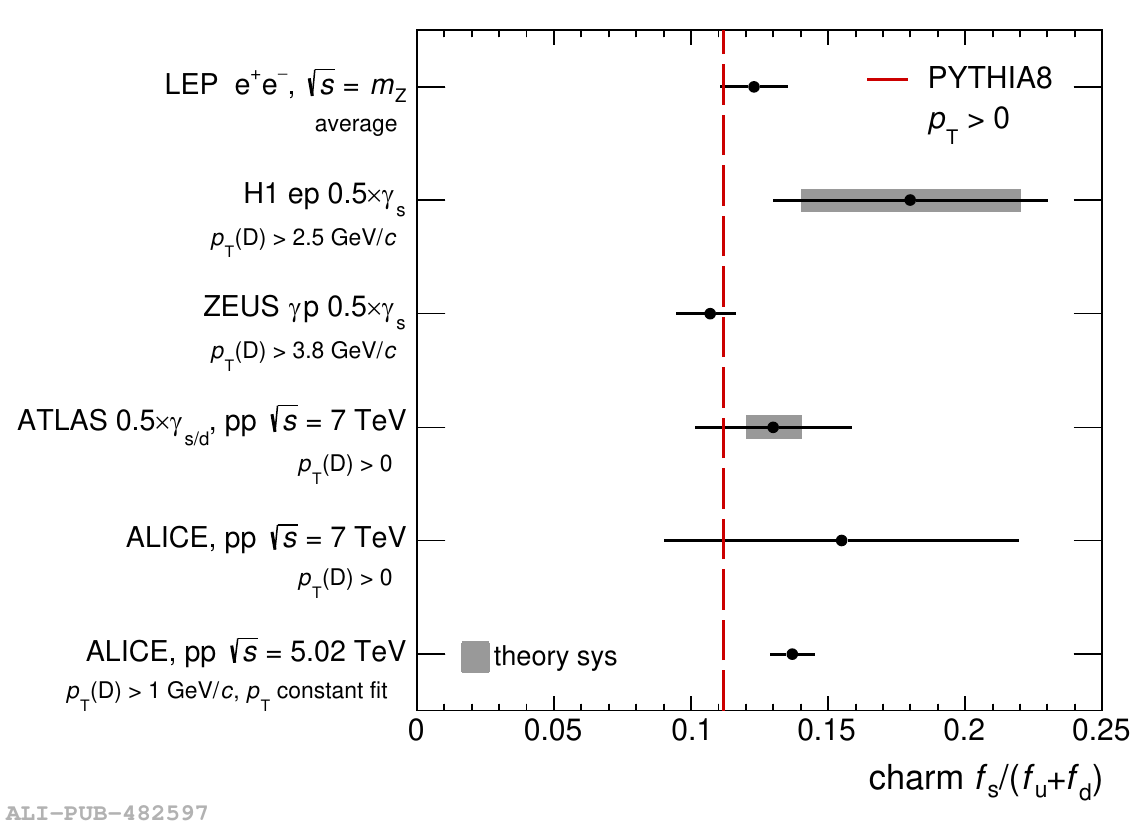}}
  \resizebox{0.45\textwidth}{!}{%
    \includegraphics{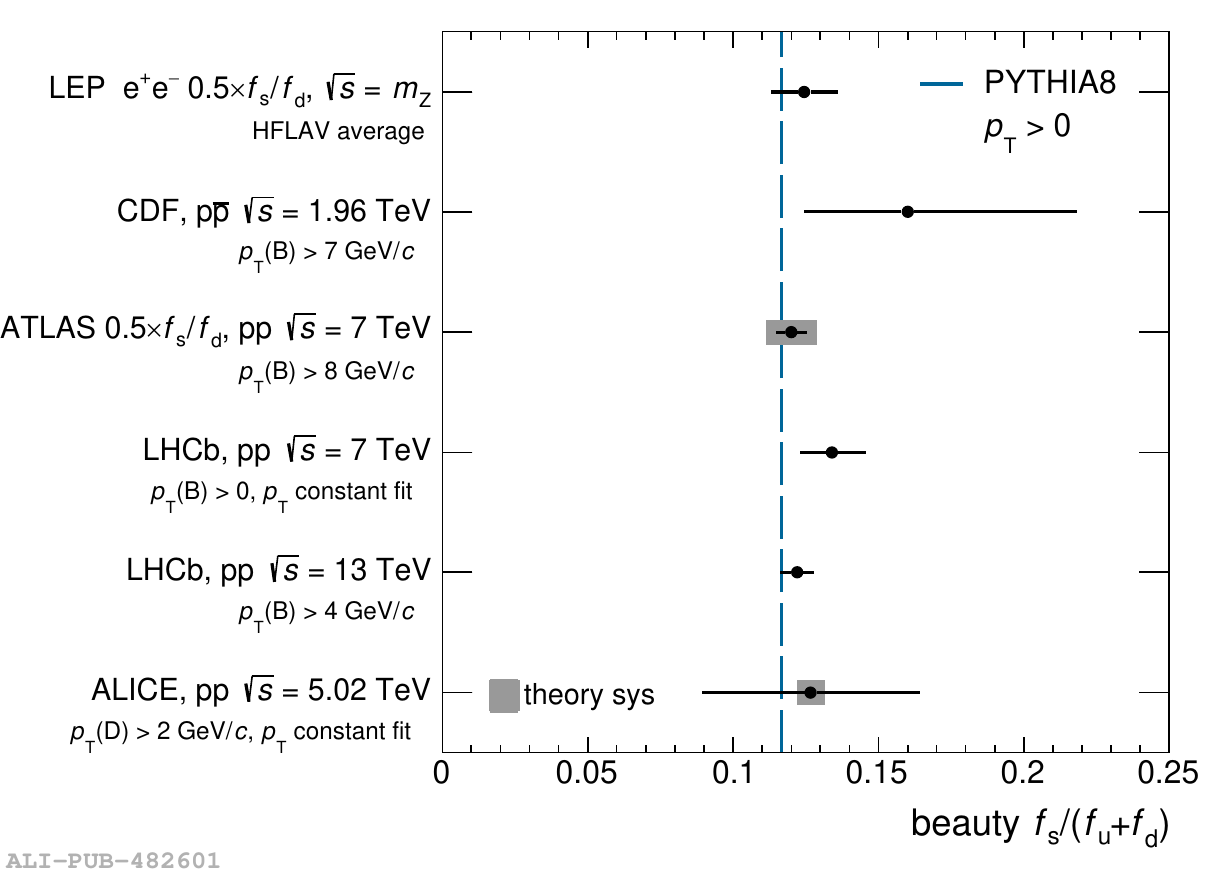}}
  \vskip-0.2cm
  \caption{(Left) Charm-quark fragmentation-fraction ratio $f_s/(f_u + f_d)$~\cite{ALICE:2021mgk}
    compared with previous
    measurements performed by the ALICE~\cite{ALICE:2012gkr},
    H1~\cite{H1:2004bwe}, ZEUS~\cite{ZEUS:2013fws},
    and ATLAS~\cite{ATLAS:2015igt} collaborations and to the average
    of LEP measurements~\cite{Gladilin:2014tba}.
    (Right) Bottom-quark fragmentation-fraction ratio $f_s/(f_u + f_d)$~\cite{ALICE:2021mgk}
    compared with previous
    measurements performed
    by the CDF~\cite{CDF:2008yux},
    LHCb~\cite{LHCb-PAPER-2011-018,LHCb-PAPER-2018-050}, and
    ATLAS~\cite{ATLAS:2015esn} collaborations and to
    the average of LEP measurements~\cite{HFLAV:2019otj}.}
   \label{fig:AlicefsVsfd}
 \end{figure}

Production cross-sections of prompt $\Dz, \Dp, \Ds$
mesons and those from $b$-decay (non-prompt)
in $pp$ collisions at $\sqs=5.02\tev$ are measured by the ALICE
collaboration~\cite{ALICE:2021mgk}. The results are found to agree with
predictions by FONLL
calculations~\cite{Cacciari:1998it,Cacciari:2001td}, as shown in
Fig.~\ref{fig:AliceDVsFONLL}.
Figure~\ref{fig:AlicefsVsfd} shows comparisons of both the charm-quark and beauty-quark fragmentation-fraction ratio
$f_s/(f_u+f_d)$ with previous measurements~\cite{ALICE:2012gkr, H1:2004bwe, ZEUS:2013fws, ATLAS:2015igt,
  Gladilin:2014tba, CDF:2008yux,
  LHCb-PAPER-2011-018,LHCb-PAPER-2018-050,
  ATLAS:2015esn, HFLAV:2019otj},
and no significant dependence on centre-of-mass energy and
collision system is seen within the current precision.

\begin{figure}
  \centering
  \resizebox{0.4\textwidth}{!}{%
    \includegraphics{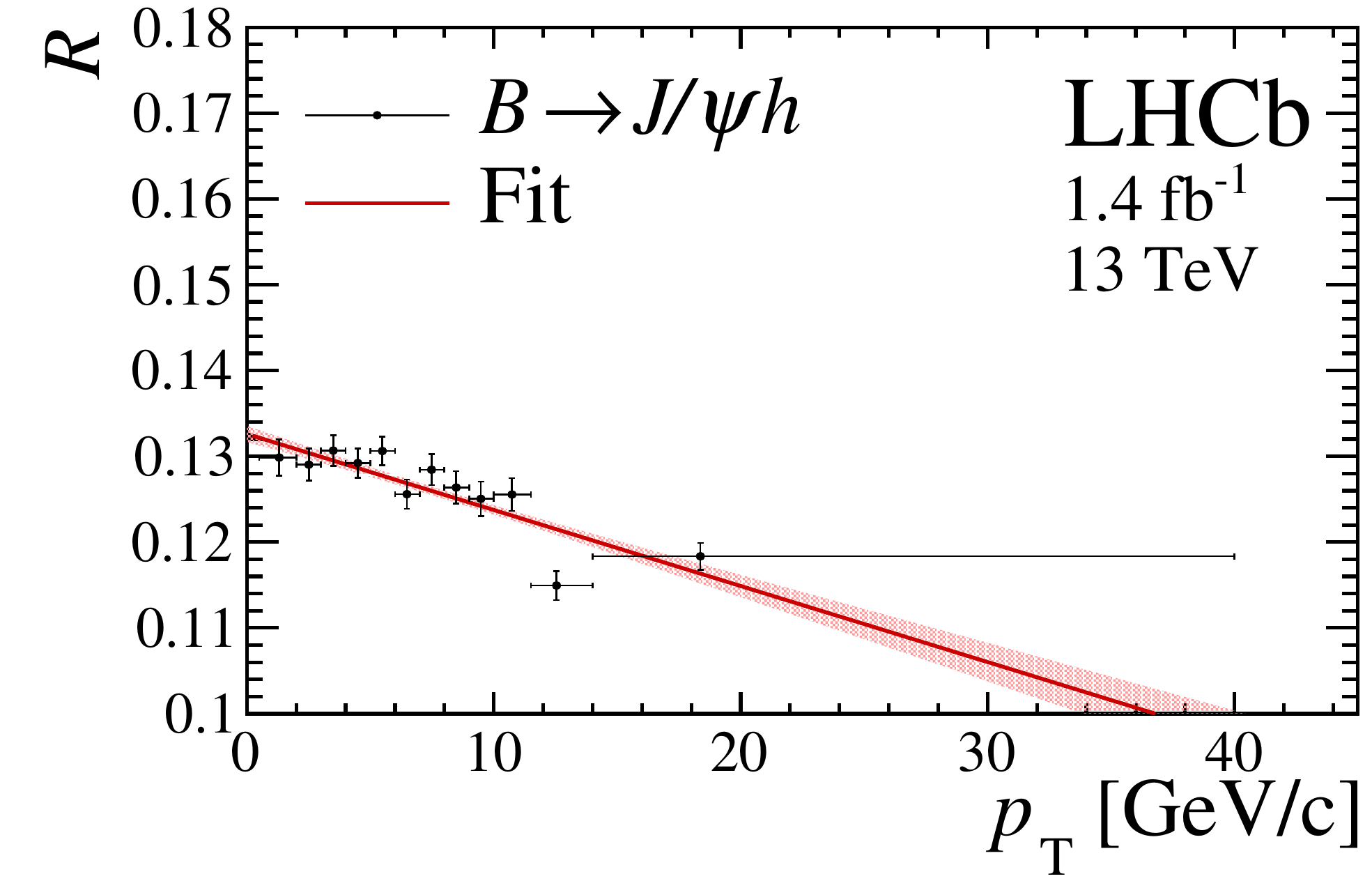}}
  \resizebox{0.4\textwidth}{!}{%
    \includegraphics{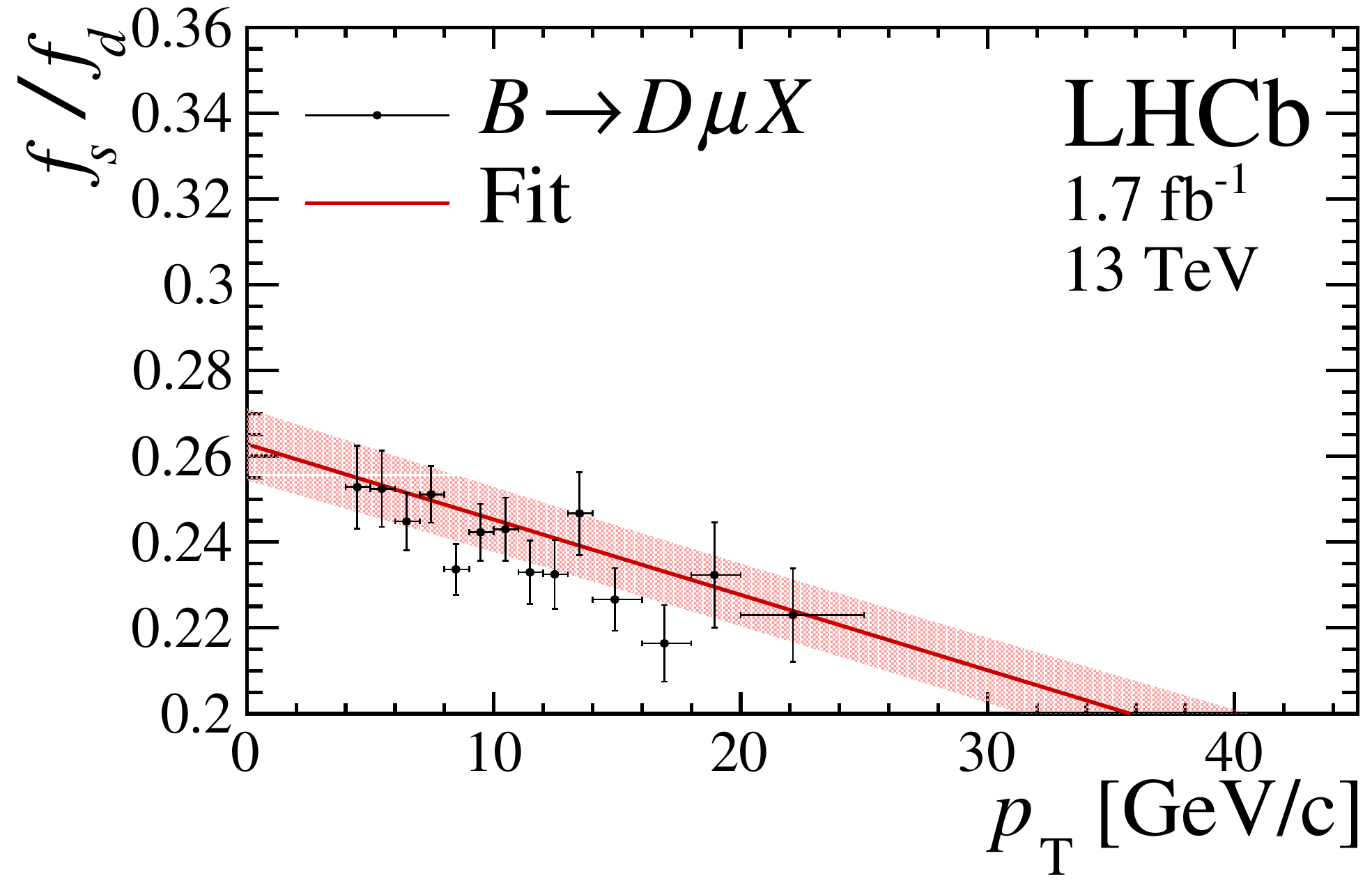}}
  \vskip-0.2cm
  \caption{Ratio of the beauty-quark fragmentation-fraction,
    $f_s/f_d$ in $pp$ collisions at $\sqs=13$ TeV, as a function of the
    $B$ transverse momentum~\cite{LHCb-PAPER-2020-046}.}
  \label{fig:LHCbfsfd}
\vskip-0.4cm
\end{figure}

The beauty-quark fragmentation-fraction ratio, $f_s/f_d$ in $pp$ collisions
 is obtained from a combined analysis~\cite{LHCb-PAPER-2020-046} of different $B$-decay modes
 measured by the LHCb experiment.
The ratio $f_s/f_d$ has clear dependence on the transverse momentum, as
shown in Fig.~\ref{fig:LHCbfsfd}, while there is no evident dependence
on the collision centre-of-mass energy. Branching fractions of $\Bs$
meson are also updated.

\begin{figure}
    \vskip-0.8cm
    \centering
  \resizebox{0.41\textwidth}{!}{%
    \includegraphics{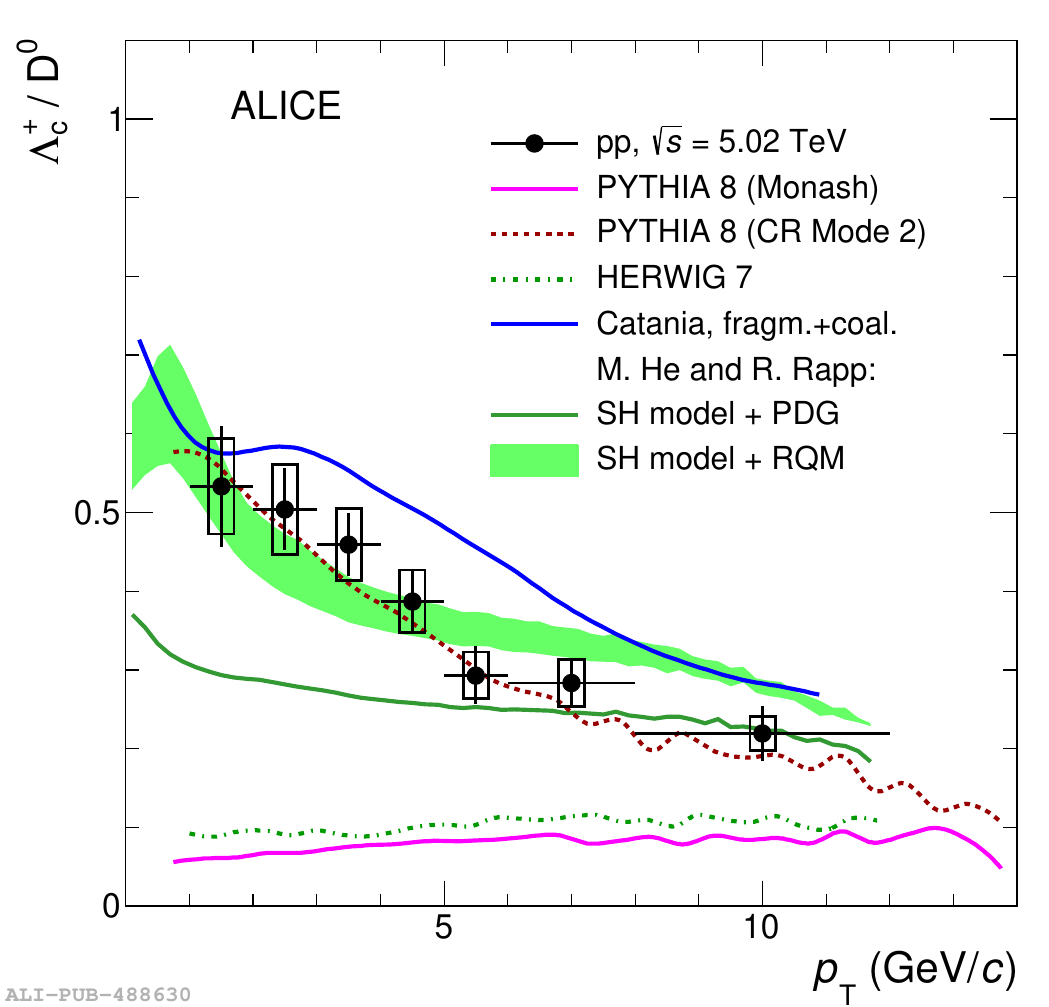}}
  \resizebox{0.45\textwidth}{!}{%
    \includegraphics{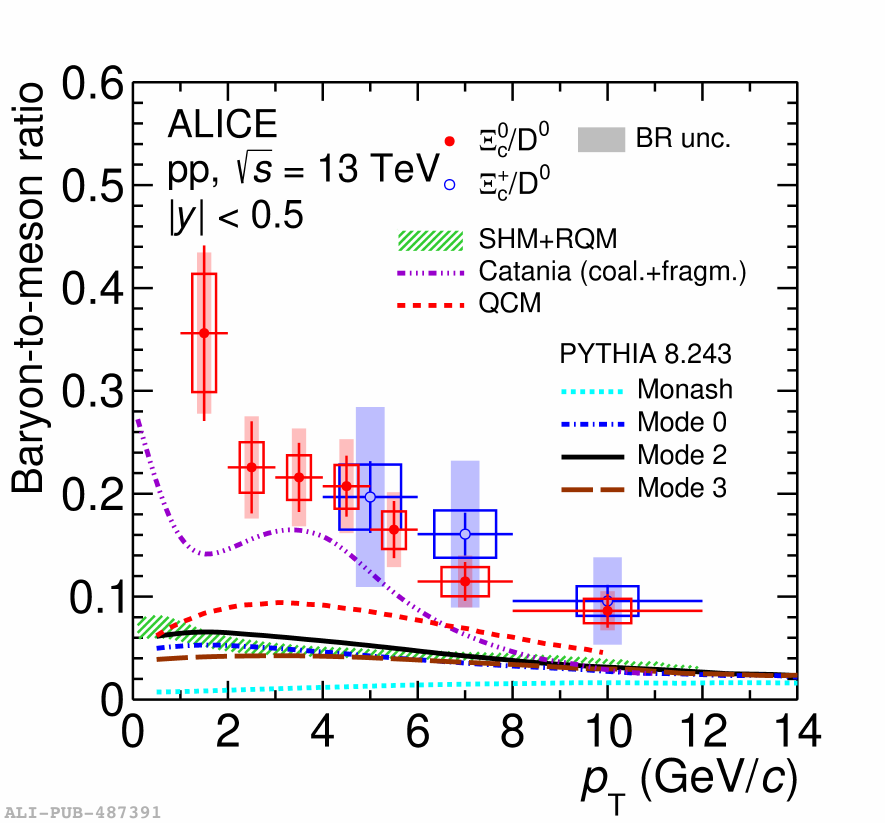}}
  \vskip-0.2cm
  \caption{Production cross-section ratio of (left) $\Lc/\Dz$ and
    (right) $\Xic/\Dz$, as function of the transverse momentum~\cite{ALICE:2020wfu,ALICE:2021bli}.}
   \label{fig:AliceHcVsD}
 \end{figure}

 Production cross-sections of charmed baryons in $pp$ collisions
 are measured by the ALICE collaboration~\cite{ALICE:2020wfu,ALICE:2021bli}.
 The production cross-section ratios of $\Lc/\Dz$ and $\Xic/\Dz$
 have clear
 dependence on the transverse momentum, as shown in
 Fig.~\ref{fig:AliceHcVsD}.
 The trend of the $\Lc/\Dz$ production ratio is 
 well decribed by
\pythia8 with a new Colour-Reconnection (CR)
mode~\cite{Christiansen:2015yqa}, 
the statistical Hadronisation (SH) including additional excited charm
baryons predicted by Relativistic Quark Model (RQM)~\cite{He:2019tik},
and the Catania model assuming that the hadronisation can occur via
coalescence in addition to fragmentation~\cite{Minissale:2020bif}.
However, none of the theoretical models can describe the trend of the $\Xic/\Dz$
production ratio yet.
Such trend of the baryon-to-meson production ratio was also seen in
beauty system by
the LHCb experiment, using both the semileptonic
decays~\cite{LHCb-PAPER-2011-018}
and fully reconstructed decays~\cite{LHCb-PAPER-2014-004,LHCb-PAPER-2015-032}. 

\begin{figure}
   \vskip-0.2cm
  \centering
  \resizebox{0.42\textwidth}{!}{%
    \includegraphics{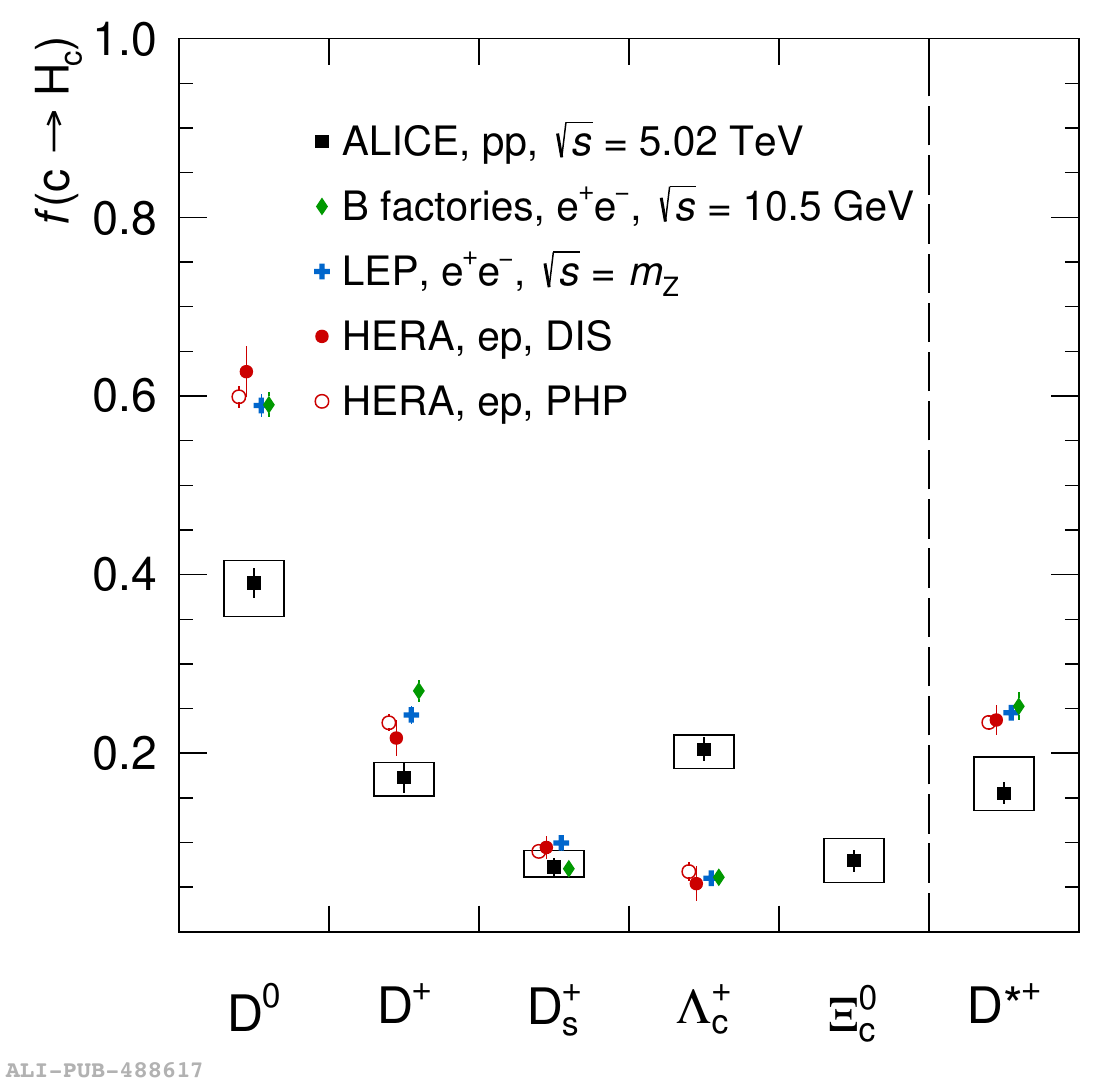}}
  \resizebox{0.42\textwidth}{!}{%
    \includegraphics{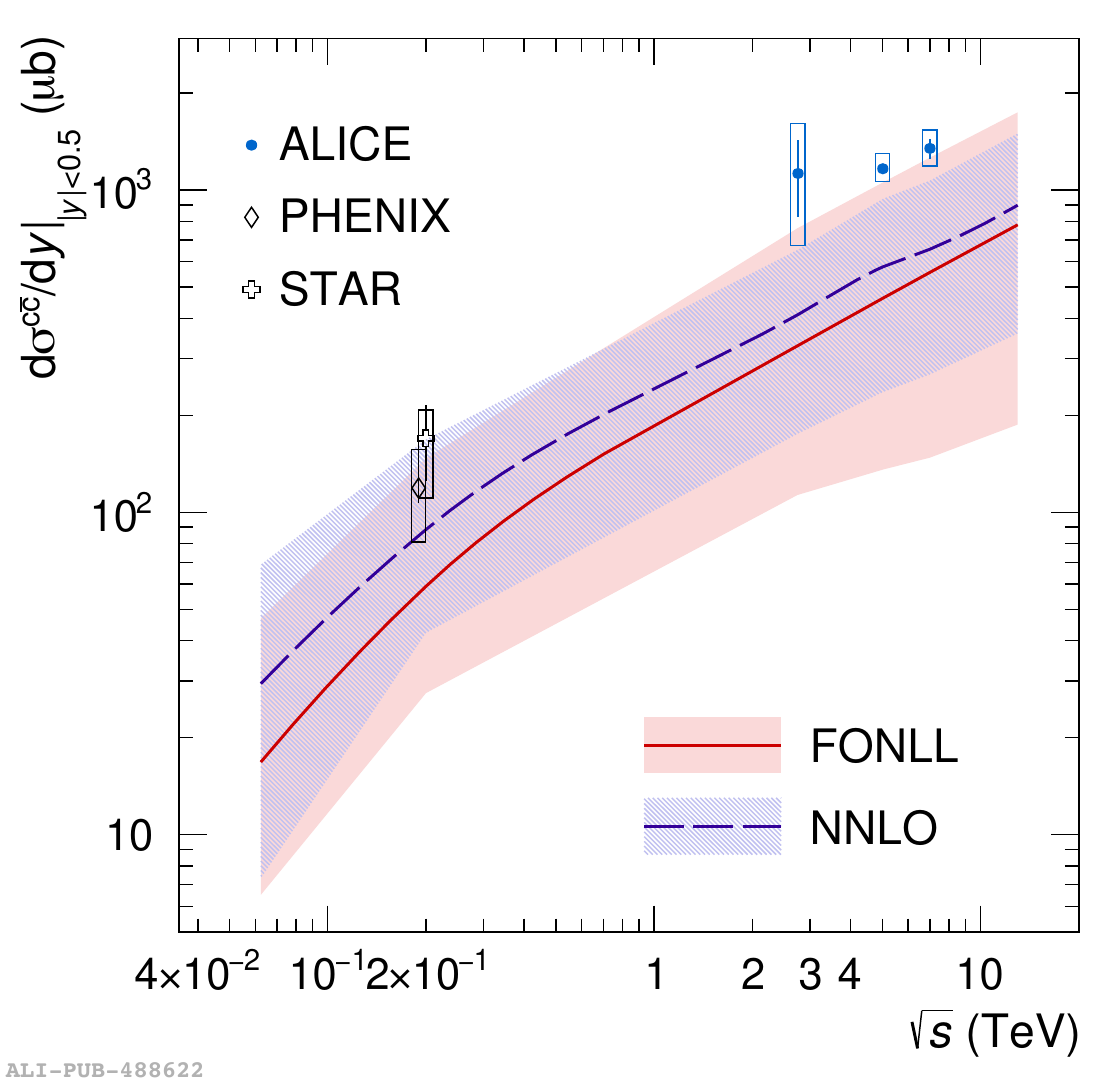}}
  \vskip-0.2cm
  \caption{(Left) Charm-quark fragmentation fractions into charm
    hadrons measured in $pp$ collisions at $\sqs= 5.02$ TeV~\cite{ALICE:2021dhb} in comparison
    with experimental measurements performed in $e^+e^-$ collisions at LEP
    and at B factories, and in $ep$ collisions at
    HERA~\cite{Lisovyi:2015uqa}.
    (Right) Charm production cross-section at midrapidity per unit of
    rapidity as a function of the collision
    energy~\cite{ALICE:2021dhb,STAR:2012nbd,PHENIX:2010xji}, and
    comparison with
    FONLL~\cite{Cacciari:1998it,Cacciari:2001td} 
    and NNLO~\cite{dEnterria:2016ids} calculations.  
  }
   \label{fig:Alice_c2HcComp}
 \end{figure}

Charm-quark fragmentation fractions are measured by the ALICE
collaboration~\cite{ALICE:2021dhb} and are compared with previous
measurements in other collision systems, as shown in
Fig.~\ref{fig:Alice_c2HcComp}. One can see that the
fragmentation fractions in $pp$ collisions differ significantly from
those obtained in
the $e^+e^-$ or $ep$ collision system when the baryons are included.
Figure~\ref{fig:Alice_c2HcComp} also shows charm production
cross-section at midrapidity
as a function of the collision
energy~\cite{ALICE:2021dhb,STAR:2012nbd,PHENIX:2010xji}, which is at
the upper edge of the
FONLL~\cite{Cacciari:1998it,Cacciari:2001td} 
and NNLO~\cite{dEnterria:2016ids} calculations.
    
\begin{figure}
  \vskip-0.5cm
  \centering
  \resizebox{0.45\textwidth}{!}{%
    \includegraphics{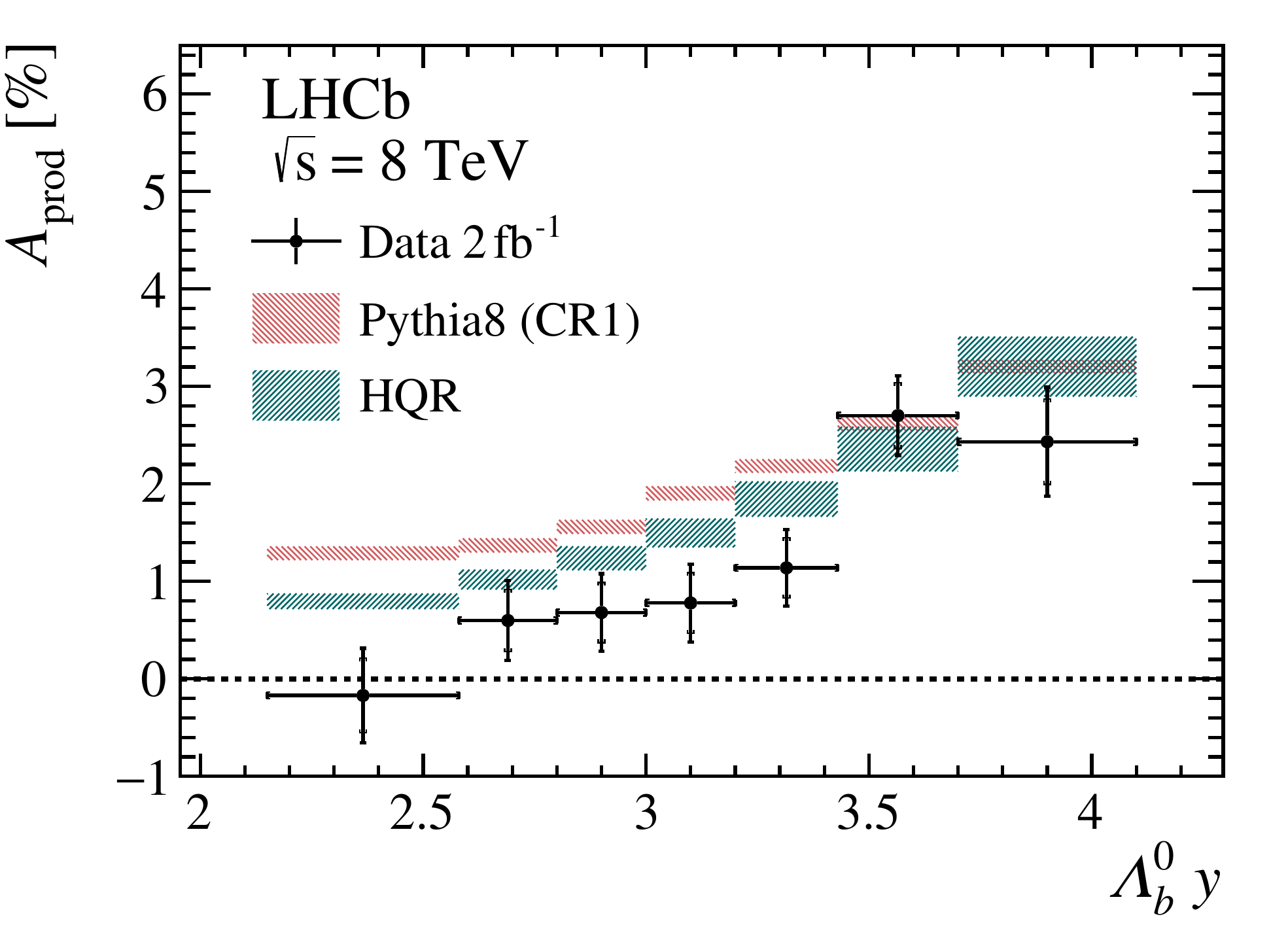}}
  \resizebox{0.45\textwidth}{!}{%
    \includegraphics{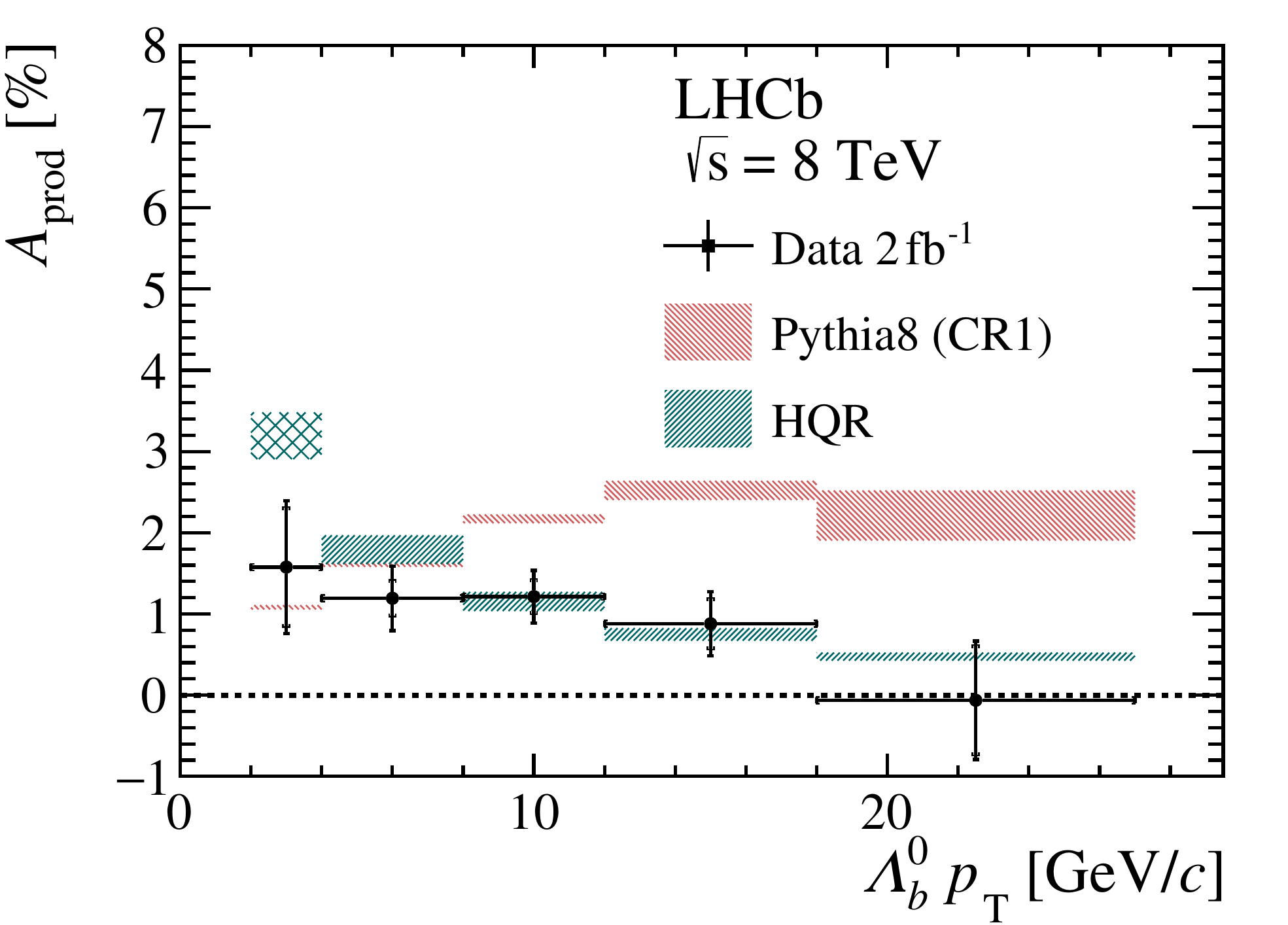}}
  \vskip-0.2cm
  \caption{Comparison of the measured $\Lb$ production asymmetry~\cite{LHCb-PAPER-2021-016}
    with the predictions by the most compatible Pythia model~\cite{Christiansen:2015yqa,Argyropoulos:2014zoa} and the heavy-quark recombination
    model (HQR)~\cite{Lai:2014iji}.}
   \label{fig:LbAsym}
\end{figure}

Differential measurements of the asymmetry between $\Lb$ and $\Lbbar$
baryon production rates in $pp$ collisions at centre-of-mass
energies of $\sqs=7$ and $8\tev$ are performed by the
LHCb collaboration~\cite{LHCb-PAPER-2021-016} using the inclusive semileptonic decay
$\Lb\to\Lc\mun\neumb X$.
The measured production asymmetry as functions of $\Lb$ rapidity $y$ and
transverse momentum $\pt$,
and comparisons 
to predictions from hadronisation models in
\pythia~\cite{Christiansen:2015yqa,Argyropoulos:2014zoa}
and heavy-quark recombination~\cite{Lai:2014iji},
are shown in Fig.~\ref{fig:LbAsym}, which shows a moderate agreement.
This is the first observation of a
particle-antiparticle asymmetry in $b$-hadron production at LHC
energies.

The $B_c^{(*)}(2S)^{+}$ states were observed by the CMS and LHCb
collaborations~\cite{CMS:2019uhm,LHCb-PAPER-2019-007},
and their production rates were measured by the CMS
collaboration~\cite{CMS:2020rcj}.
The $B_c(2S)^+$ to $\Bc$, $B_c^*(2S)^+$ to \Bc, and $B_c^*(2S)^+$ to
$B_c(2S)^+$ cross-section ratios,  including the unknown
$B_c^{(*)}(2S)^+\to B_c^{(*)+}\pip\pim$
branching fractions, are determined to be
$(3.47 \pm 0.63  \pm 0.33 )\%$, $(4.69 \pm 0.71 \pm 0.56)\%$, and  $1.35 \pm 0.32 \pm 0.09$, respectively.
They have no significant dependence on the $\Bc$ \pt and $y$. 

\section{Heavy flavour spectroscopy}
\vskip-0.2cm

\begin{figure}
  \vskip-0.4cm
\begin{minipage}{0.5\textwidth}
  \centering
  \resizebox{0.75\textwidth}{!}{    
    \includegraphics{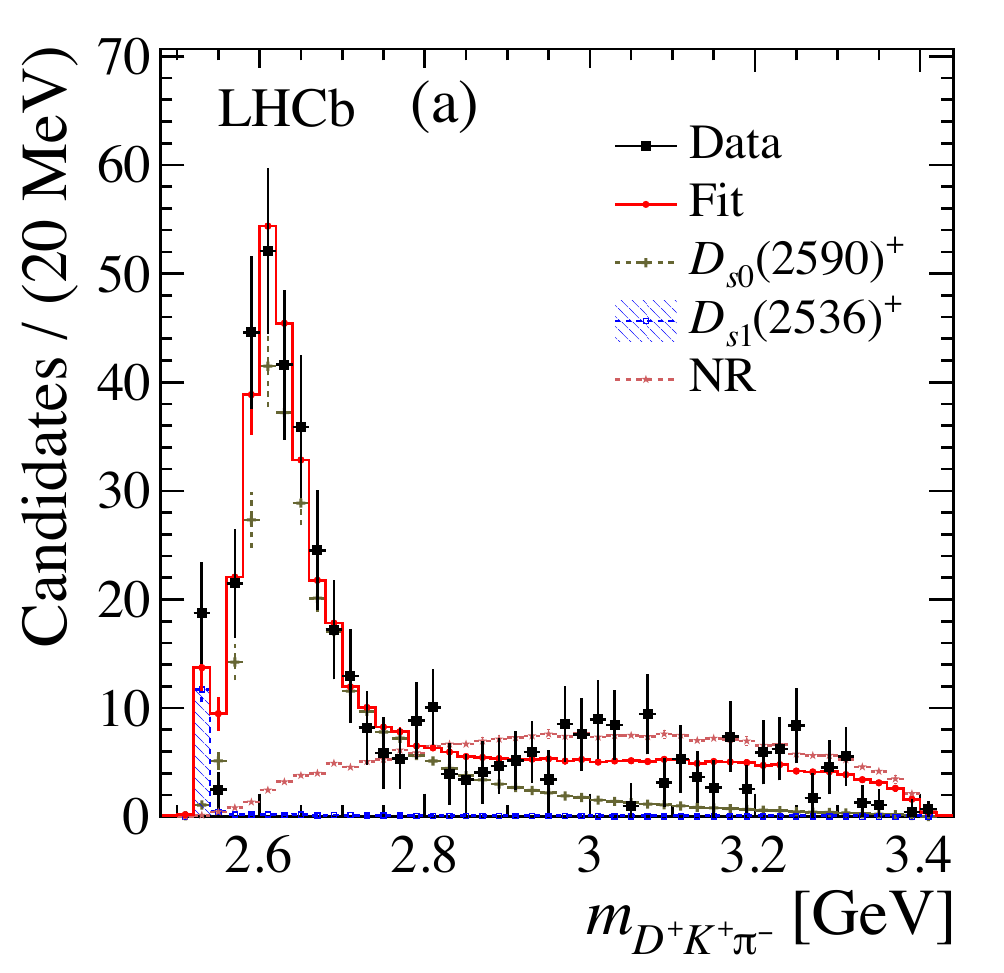}}
  \vskip-0.2cm
   \caption{Projection of $m(D^+K^+\pi^-)$
     in the amplitude analysis of the $\Bd\to D^-D^+K^+\pi^-$ decay~\cite{LHCb-PAPER-2020-034}. 
   }
   \label{fig:Ds2590}
\end{minipage}\hfill
\begin{minipage}{0.48\textwidth}
  \centering
  \resizebox{0.95\textwidth}{!}{    
    \includegraphics{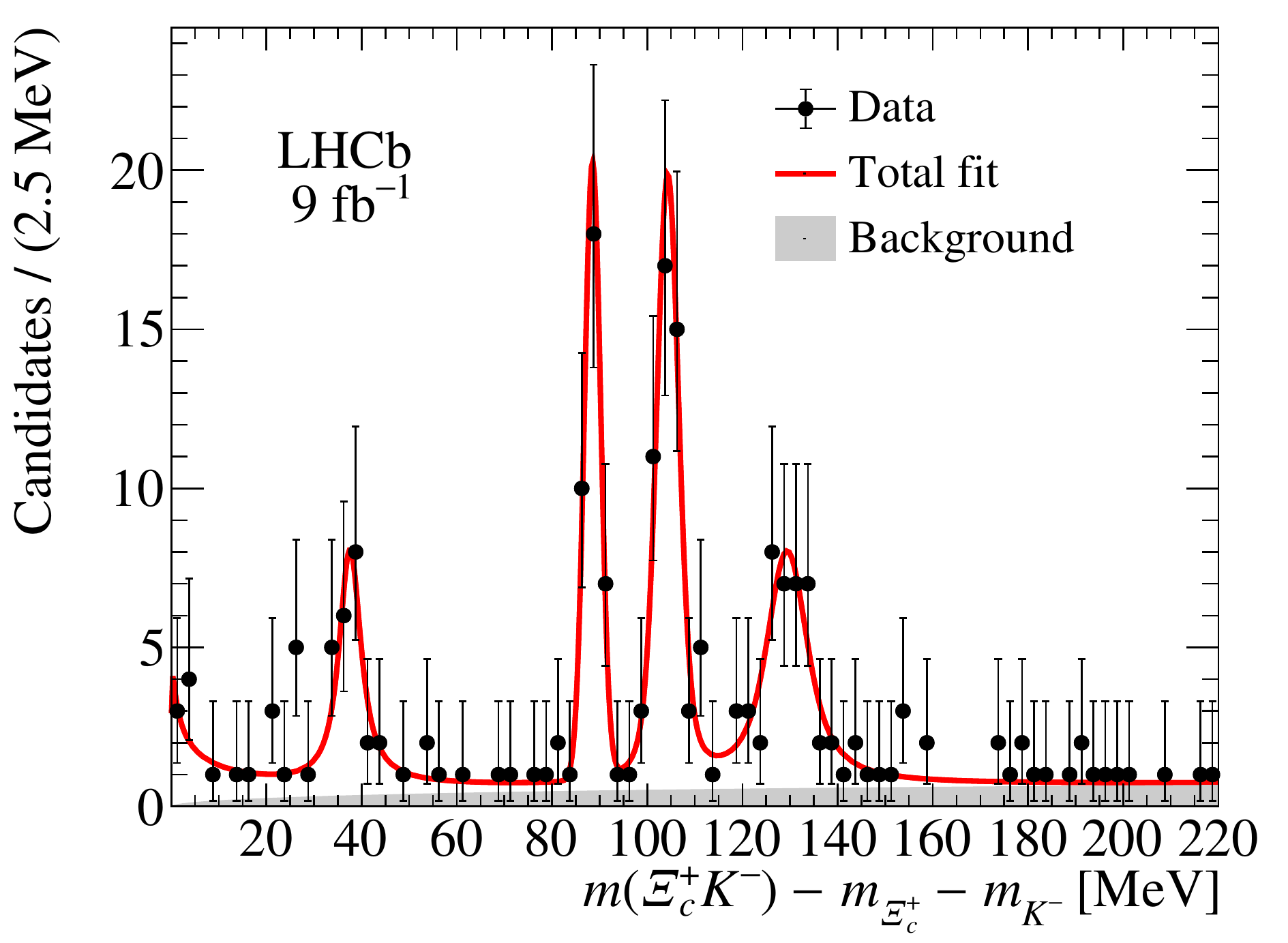}}
  \vskip-0.3cm
   \caption{Distribution of the reconstructed mass difference between
     $m(\Xicp\Km)$ and the $\Xicp$ and $\Km$ masses~\cite{LHCb-PAPER-2021-012}. 
   }
   \label{fig:OmegacFromb}
\end{minipage}
\vskip-0.2cm
\end{figure}

In the study of the $\Bd\to D^-D^+K^+\pi^-$ decay, a new excited $\Dsp$ state
decaying to $D^+K^+\pi^-$ is observed by the LHCb
collaboration~\cite{LHCb-PAPER-2020-034}, as shown in
Fig.~\ref{fig:Ds2590}.
The pole mass and width, and the spin-parity of the new state are
measured with an amplitude analysis to be $m_R=2591\pm6\pm 7\mev$,
$\Gamma_R = 89\pm 16 \pm 12\mev$, and $J^P=0^-$.

In the study of the $\Omegab\to\Xicp\Km\pim$ decay, 
four excited $\Omegac$ baryons,
$\POmega_c(3000)^0$, $\POmega_c(3050)^0$, $\POmega_c(3065)^0$,
$\POmega_c(3090)^0$,
are observed in the $\Xicp\Km$ mass
projection~\cite{LHCb-PAPER-2021-012}, and 
there is a structure at threshold with a significance of $4.3\sigma$,
as shown in Fig.~\ref{fig:OmegacFromb}.
A test of spin hypotheses is performed,
the combined hypothesis of the four peaks to have quantum numbers in
the order $(1/2, 1/2, 3/2, 3/2)$ is tested and rejected with a
significance of $3.5\sigma$.
These four states, in addition to 
the $\POmega_c(3119)^0$ state that is not seen in this analysis,  
were observed previously in the prompt $\Xicp\Km$ mass
spectrum~\cite{LHCb-PAPER-2017-002}.

\begin{figure}
\begin{minipage}{0.50\textwidth}
  \centering
  \resizebox{0.9\textwidth}{!}{    
     \includegraphics{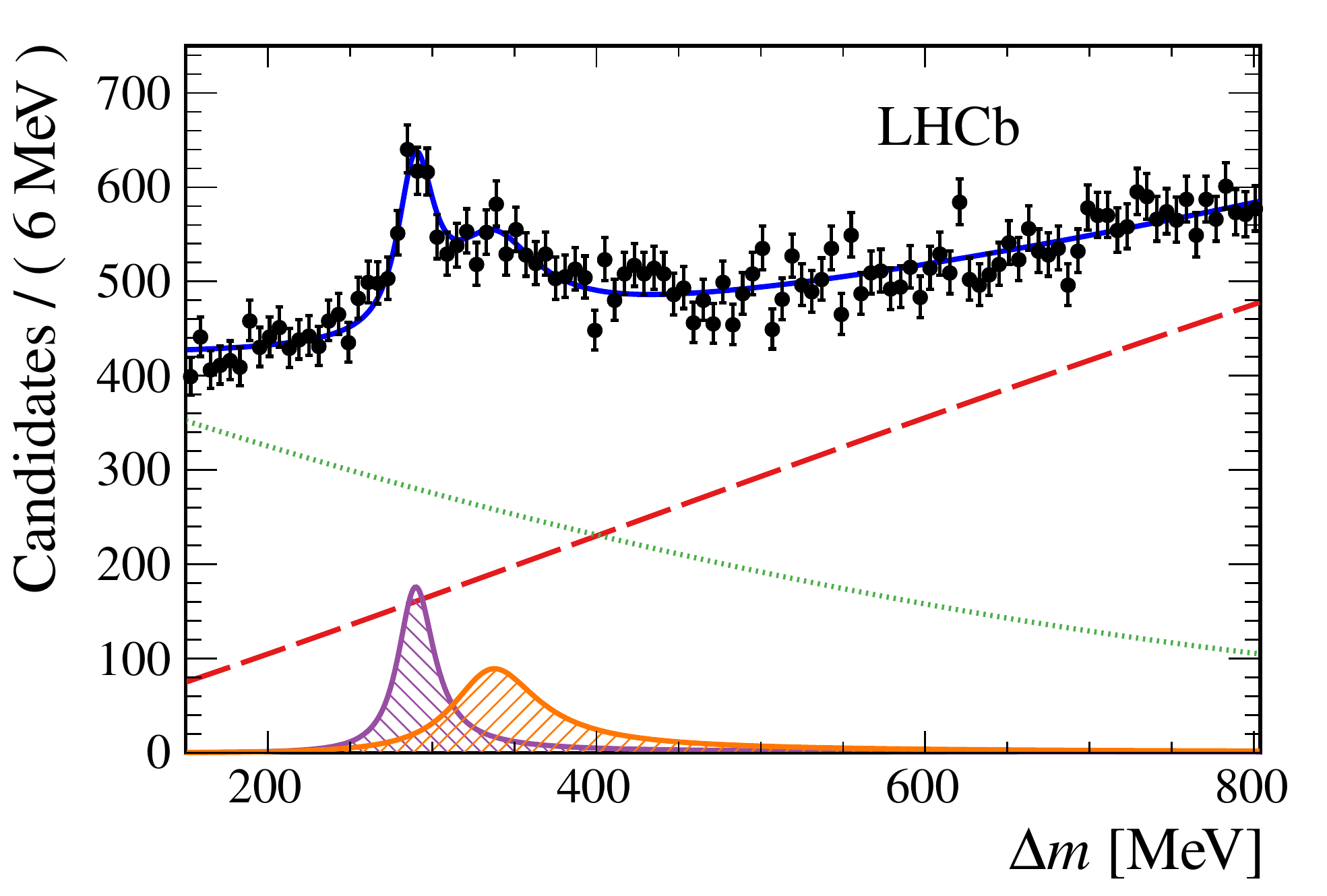}}
   \caption{Distribution of the reconstructed mass difference between
     the $m(\Bu\Km)$ and the $\Bu$ and $\Km$
     masses~\cite{LHCb-PAPER-2020-026}.
     Associated production (combinatorial) is shown as the green
     dotted (red dashed) line.
   }
   \label{fig:ExcitedBs}
\end{minipage}\hfill
\begin{minipage}{0.47\textwidth}
  \centering
  \resizebox{0.95\textwidth}{!}{    
    \includegraphics{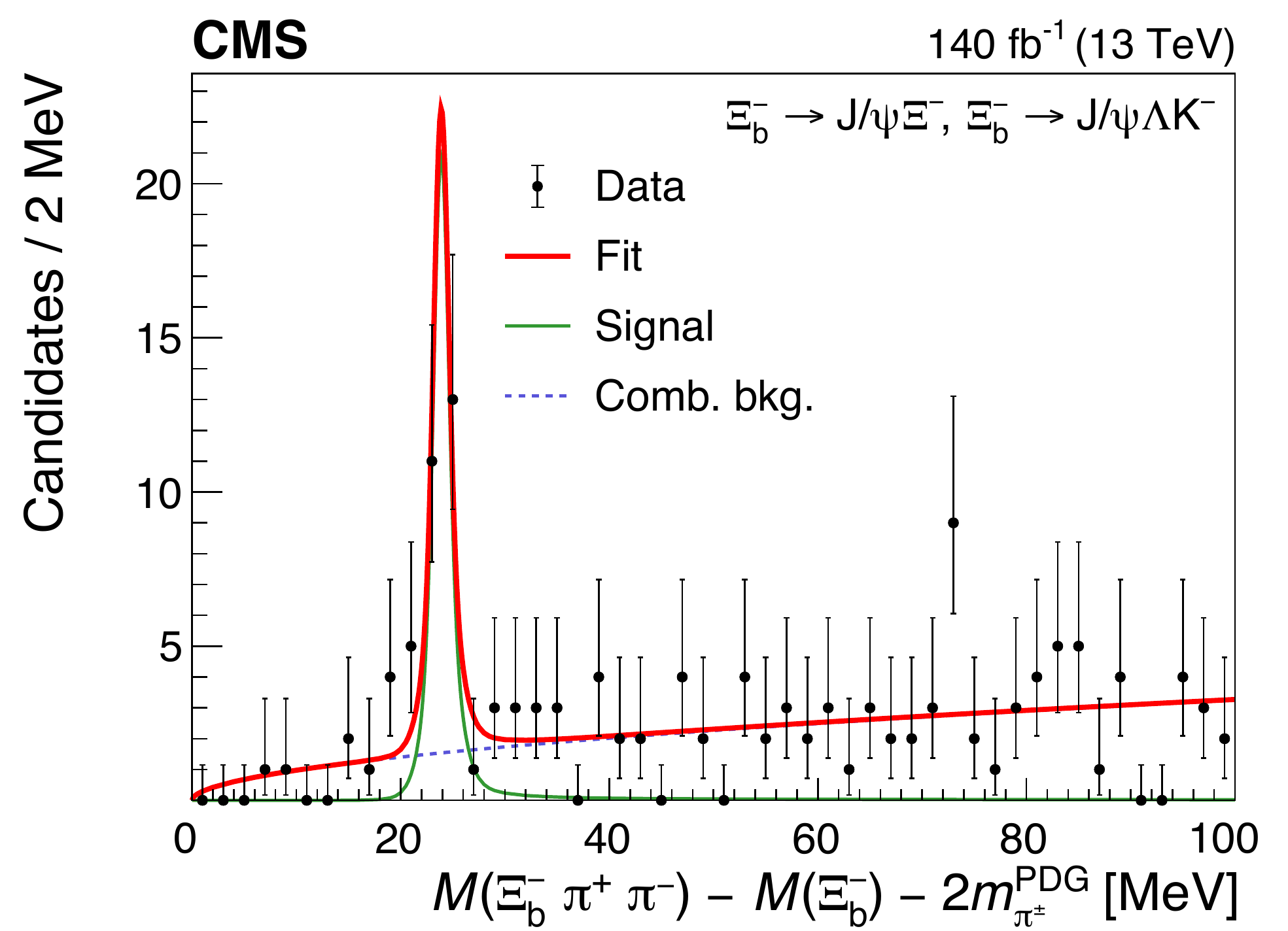}}
   \caption{Distribution of the reconstructed mass difference between
     the $\Xibm\pip\pim$ invariant mass and the $\Xibm$ and $2\pi$ masses~\cite{CMS:2021rvl}.
   }
   \label{fig:CMS_Xib6100}
\end{minipage}
\vskip-0.3cm
\end{figure}

A structure, interpreted as the result of overlapping excited $\Bs$
states, is observed in the $\Bu\Km$ mass spectrum by the LHCb
collaboration~\cite{LHCb-PAPER-2020-026}, as shown in
Fig.~\ref{fig:ExcitedBs}.
Assuming they decay directly to $\B^{+}\kaon^{-}$,
their masses and widths are determined to be:
$m_1      = 6063.5 \pm   1.2  \pm   0.8 \mev,
\Gamma_1  =  26 \pm   4  \pm   4 \mev, 
m_2       = 6114 \pm   3 \pm   5 \mev, 
\Gamma_2  =  66 \pm   18 \pm   21 \mev.$
Alternative values assuming a decay through $\B^{*+}\kaon^{-}$,
with a missing photon from the $B^{*+} \to B^{+}\gamma$ decay,
would be shifted by approximately 45\mev. 

\begin{figure}[t]
  \vskip-0.8cm
  \begin{minipage}{0.46\textwidth}    
  \centering
  \resizebox{0.7\textwidth}{!}{%
    \includegraphics{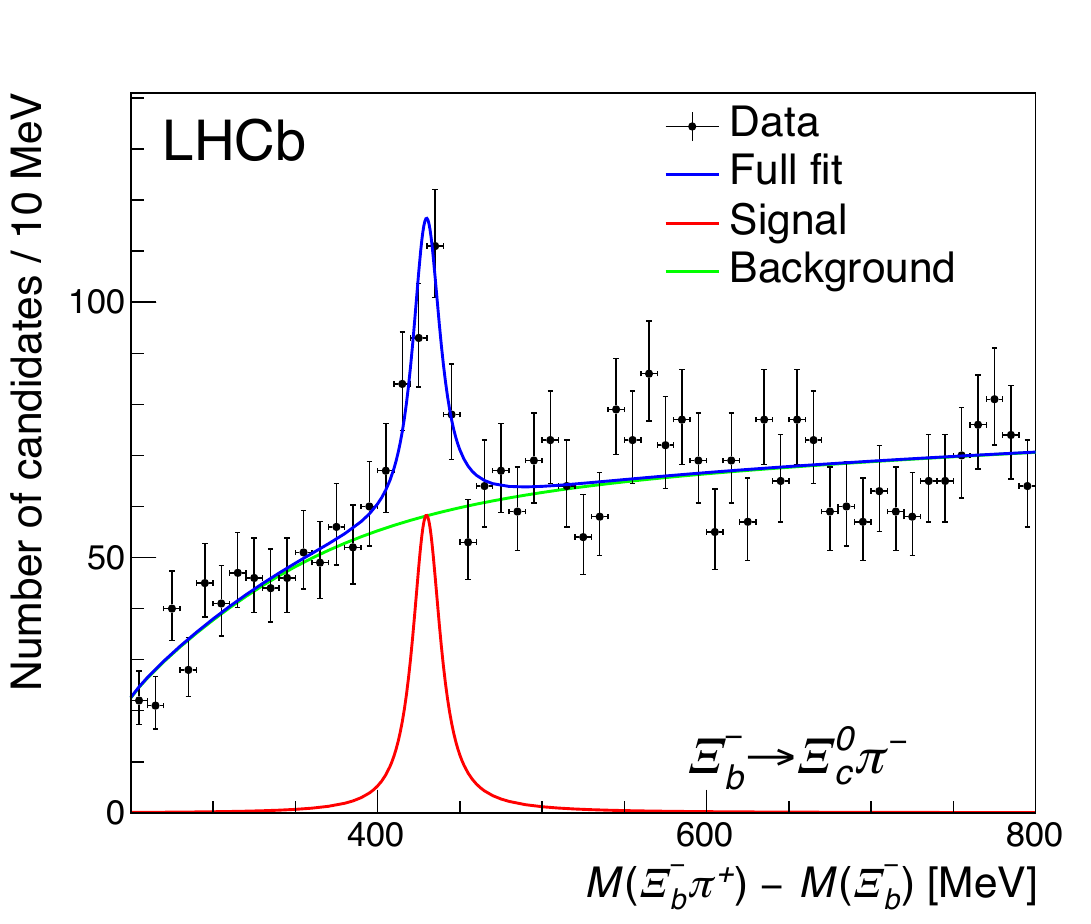}}
  \vskip-0.2cm
  \caption{Distribution of the reconstructed mass difference between
     the $m(\Xibm\pip)$ and the $\Xibm$ and $\pip$ masses~\cite{LHCb-PAPER-2020-032}.}
   \label{fig:Xib6227}
 \end{minipage}\hfill
 \begin{minipage}{0.5\textwidth}
  \centering
   \begin{overpic}[scale=1.0]{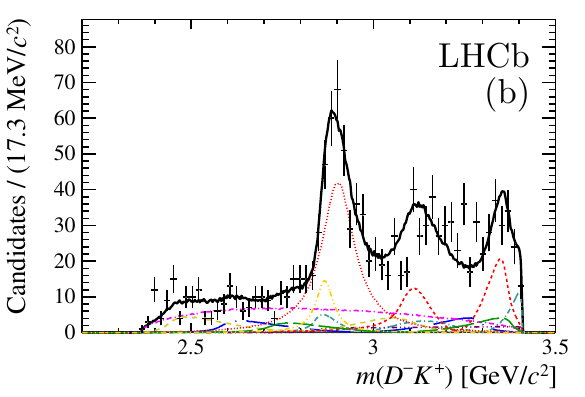}
     \put(15,35){\includegraphics[scale=0.4]{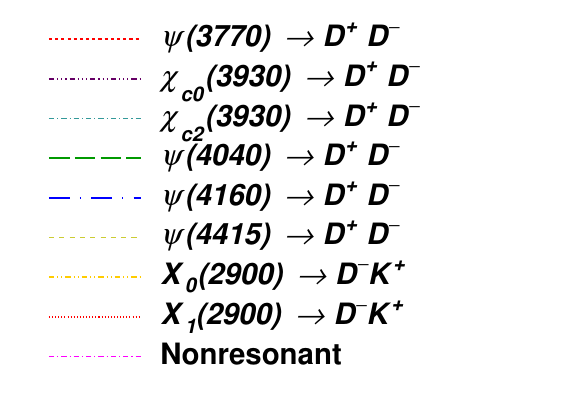}}
   \end{overpic}
   \vskip-0.2cm
   \caption{Projection of $m(D^-K^+)$ in the amplitude analysis of the
     $B^+\to D^+D^-K^+$ decay~\cite{LHCb-PAPER-2020-025}.}
   \label{fig:X2900}
 \end{minipage}
 %\end{figure}
 %\begin{figure}
 \begin{minipage}{0.5\textwidth}
  \centering
  \resizebox{1.0\textwidth}{!}{    
    \includegraphics{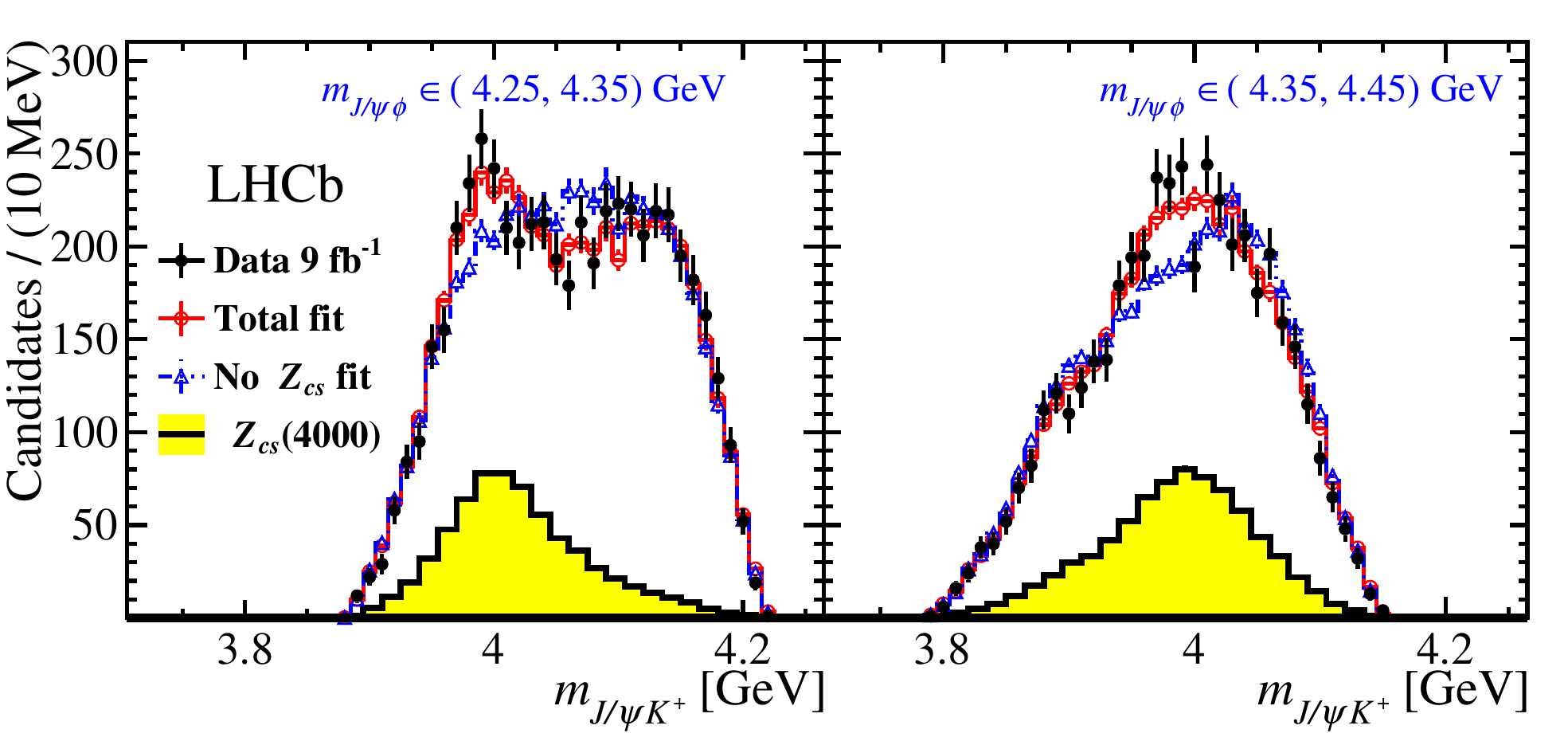}}
  \vskip-0.2cm
   \caption{Projection of $m(\jpsi K^+)$ in the amplitude analysis of the
     $B^+\to \jpsi\phi K^+$ decay in two slices of $m(\jpsi\phi)$~\cite{LHCb-PAPER-2020-044}.}
   \label{fig:Zcs}
 \end{minipage}\hfill
 \begin{minipage}{0.45\textwidth}
  \centering
  \resizebox{1.0\textwidth}{!}{    
    \includegraphics{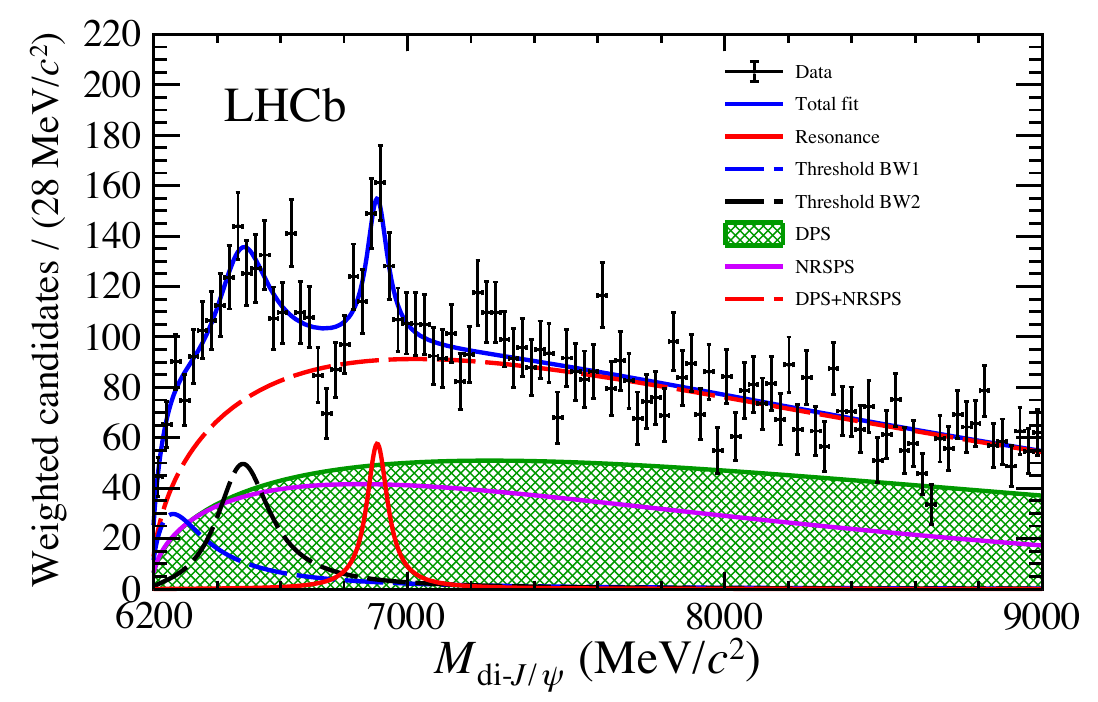}}
  \vskip-0.3cm
   \caption{Distribution of invariant mass of weighted di-$\jpsi$
     candidates~\cite{LHCb-PAPER-2020-011}.}
   \label{fig:X6900}
 \end{minipage}
%\vskip-0.2cm
 %\end{figure}
 %\begin{figure}%
 \begin{minipage}{0.45\textwidth}
  \centering
  \resizebox{0.9\textwidth}{!}{    
    \includegraphics{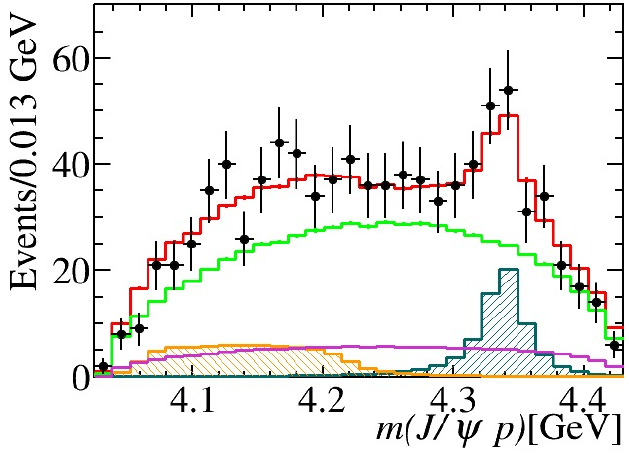}}
  \vskip-0.3cm
   \caption{Projection of $m(\jpsi p)$ in the amplitude analysis of the
     $\Bs\to \jpsi p \antiproton$ decay~\cite{LHCb-PAPER-2021-018}.}
   \label{fig:Bs2Jpsipp}
 \end{minipage}\hfill   
 \begin{minipage}{0.49\textwidth}
  \centering
  \resizebox{0.95\textwidth}{!}{    
    \includegraphics{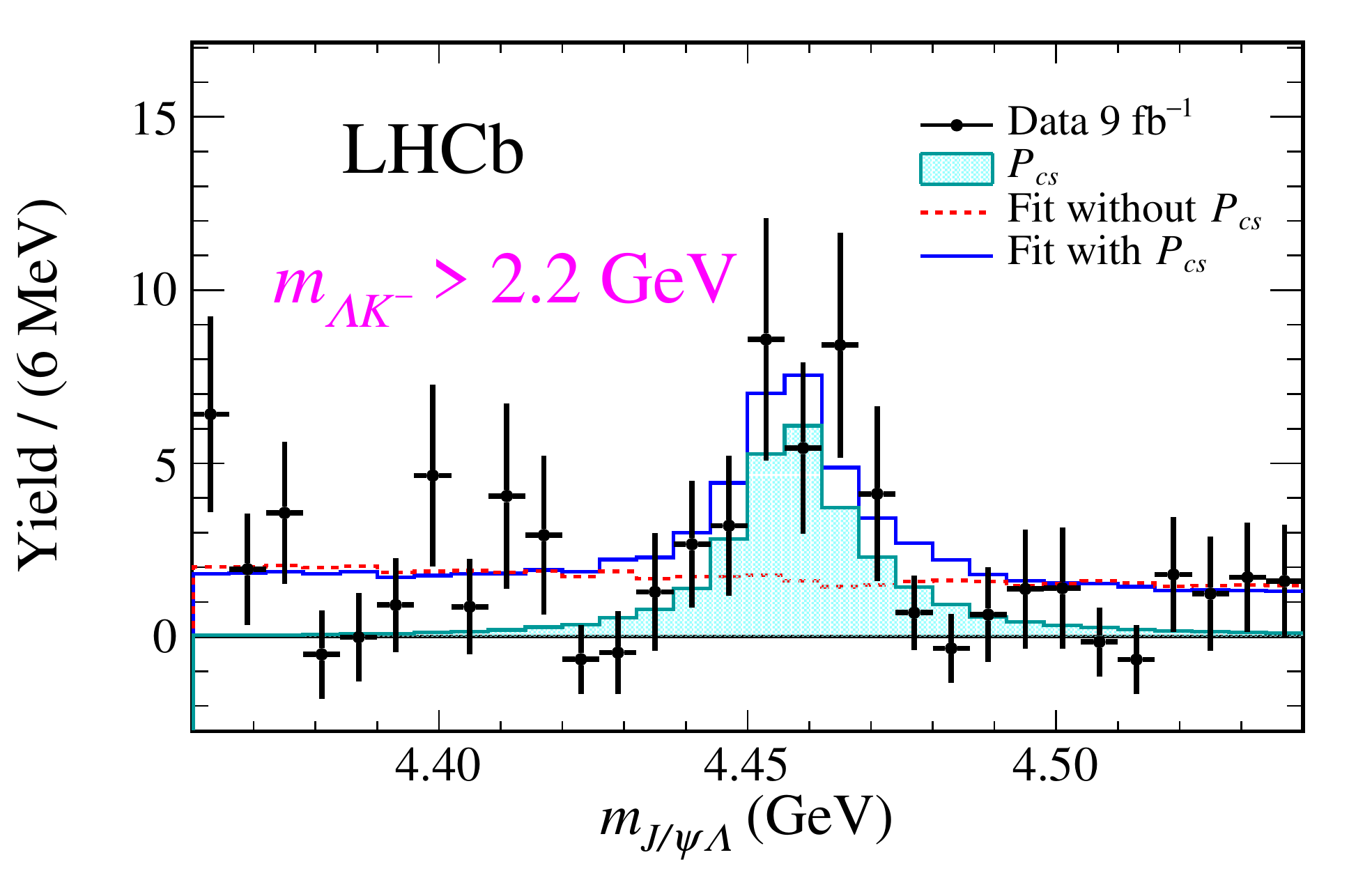}}
  \vskip-0.2cm
   \caption{Projection of $m(\jpsi \PLambda)$ in the amplitude analysis of the
     $\Xibm\to\jpsi\PLambda\Km$ decay with $m(\PLambda\Km)>2.2\gev$~\cite{LHCb-PAPER-2020-039}.}
   \label{fig:Pcs}
 \end{minipage}
 \vskip-0.3cm
 \end{figure}

For excited $b$-baryon states, 
a narrow resonance $\Xib(6100)^-$ is observed in the $\Xibm\pip\pim$
mass spectrum by the CMS collaboration~\cite{CMS:2021rvl}.
The $\Xibm$ is reconstructed via its decays to
$\jpsi\PXi^-$ and $\jpsi\PLambda\Km$, and the $\Xibm\pip\pim$
mass spectrum with the fully reconstructed $\Xibm$ is shown in
Fig.~\ref{fig:CMS_Xib6100}.
The $\Xib(6100)^-$ mass is determined to be $6100.3\pm 0.2\pm0.1\pm
0.6(\Xibm)\mev$, where the last uncertainty reflects the precision of
the $\Xibm$ baryon mass. An upper limit of 1.9\mev at 95\% CL is set on
its natural width. 
% LHCb
A new excited $\Xibz$ resonance, $\Xib(6227)^0$, is observed in the
$\Xibm\pi^+$ mass spectrum by the LHCb
collaboration~\cite{LHCb-PAPER-2020-032}.
The $\Xibm$ state is reconstructed in the fully hadronic decay modes
$\Xicz\pim$ and $\Xicz\pim\pip\pim$, and the $\Xibm\pip$ mass spectrum
with $\Xibm\to\Xicz\pim$ is shown in Fig.~\ref{fig:Xib6227}.
The mass and natural widths are determined to be
$m=6227.1^{+1.4}_{-1.5}\pm0.5\mev, \Gamma=18.6^{+5.0}_{-4.1}\pm 1.4\mev$. 
Improved measurements of the mass and natural width of the previously
observed $\Xib(6227)^-$ state~\cite{LHCb-PAPER-2018-013}, along with the mass of the $\Xibm$
baryon, are also reported~\cite{LHCb-PAPER-2020-032}.

A search for the doubly charmed baryon $\Xiccp$ is performed in the
$\Xicp\pip\pim$ decay mode by the LHCb
collaboration~\cite{LHCb-PAPER-2021-019}.
No significant signal is seen, and upper
limits on its production rate are set for different
$\Xiccp$ mass and lifetime hypotheses. The results from this search are
combined with a previously published search for the  $\Xiccp \to
\Lc\Km\pip$ decay mode~\cite{LHCb-PAPER-2019-029}, yielding a maximum local (global) significance of
$4.0\sigma\;(2.9\sigma)$ around the known mass of its isospin partner $\Xiccpp$~\cite{LHCb-PAPER-2019-037}, including
systematic uncertainties.

The LHCb collaboration performed the first search for the doubly charmed baryon with strangeness
$\POmega_{cc}^+$ via its decay to $\Xicp\Km\pip$~\cite{LHCb-PAPER-2021-011},
the doubly heavy baryon $\PXi_{bc}^0$ via its decay to $D^0pK^-$~\cite{LHCb-PAPER-2020-014},
and $\PXi_{bc}^0, \POmega_{bc}^0$ via their decays to
$\Lc(\Xicp)\pi^-$~\cite{LHCb-PAPER-2021-002}, 
no significant signal is seen yet, and upper limits on their
production rates are set for different
mass and lifetime hypotheses.

In the amplitude analysis of the $\Bu\to D^+ D^- K^+$ decay performed
by the LHCb collaboration~\cite{LHCb-PAPER-2020-025}, it is
found to be necessary to
include new spin-0 and spin-1 resonances in the $D^-K^+$ channel with masses around 2.9\gev
and a new spin-0 charmonium resonance in proximity to the spin-2
$\chi_{c2}(3930)$ state, as shown in Fig.~\ref{fig:X2900}.
This is also supported by a model-independent study~\cite{LHCB-PAPER-2020-024}.
The masses and natural widths of the two exotic states in the $D^-K^+$
channel are determined
to be, $X_0(2900): M=2866\pm7\pm 2\mev, \Gamma=57\pm12\pm4 \mev$,
$X_1(2900): M=2904\pm 5\pm 1\mev, \Gamma=110 \pm 11 \pm 4 \mev$.

 In the amplitude analysis of the $\Bu\to\jpsi\phi\Kp$ decay performed
 by the LHCb collaboration~\cite{LHCb-PAPER-2020-044}, two
 $Z_{cs}$ states are observed. The most significant one,
 $Z_{cs}(4000)^+$, as shown in Fig.~\ref{fig:Zcs},
 has a mass of $4003\pm 6^{+4}_{-14}\mev$, and a
 width of $131\pm 15\pm 26\mev$, which is ten times higher than the
 natural width of
 the $Z_{cs}(3985)^+$ state $(m=3982.5^{+1.8}_{-2.6}\pm2.1\mev,\,\Gamma=12.8^{+5.3}_{-4.4}\pm 3.0\mev)$ observed by the BESIII
 collaboration~\cite{BESIII:2020qkh}. 

In the study of the invariant mass spectrum of $\jpsi$ pairs by the
LHCb collaboration~\cite{LHCb-PAPER-2020-011}, a narrow structure
around $6.9\gev$ is observed, as shown in Fig.~\ref{fig:X6900}. Its
mass and natural width are determined to be $6905\pm11\pm7\mev$,
$80\pm19\pm33\mev$ assuming no interference with the
nonresonant SPS continuum. This could be the hadron state consisting
of four charm quarks $T_{cc\bar{c}\bar{c}}$.

In the amplitude analysis of flavour-untagged $\Bs\to\jpsi p \antiproton$
performed by the LHCb collaboration~\cite{LHCb-PAPER-2021-018},
evidence for a new structure in the $\jpsi p$ and $\jpsi\antiproton$
systems is found, as shown in Fig.~\ref{fig:Bs2Jpsipp}.
Its mass and width are determined to be
$4337^{+7}_{-1}\,^{+2}_{-1}\mev$, $29^{+26}_{-12}\,^{+14}_{-14}\mev$,
respectively. 
In the amplitude analysis of the $\Xibm\to\jpsi\PLambda\Km$ decay
performed by the LHCb collaboration~\cite{LHCb-PAPER-2020-039}, 
first evidence for a charmonium pentaquark with strangeness is found,
with a mass of $4458.8\pm2.9^{+4.7}_{-1.1}\mev$ and a width of
$17.3\pm6.5 ^{+8.0}_{-5.7}\mev$, as shown in Fig.~\ref{fig:Pcs}.

 \section{Summary}
 \vskip-0.2cm
 
Great progress has been made on studies of the heavy flavour
production and spectroscopy at the LHC. This includes measurements of productions of
$D, \Lc, \Xic, \Lb, B_c^{(*)}(2S)^+$, and observations of
$D_{s0}(2590)^+$, excited $\Xibm$ states, $X(2900)$, $Z_{cs}^+$,
$X(6900)$, and evidence of $P_c(4337)^+$ and $P_{cs}$. These results are very helpful for
understanding the strong interaction. 

\addcontentsline{toc}{section}{References}
\setboolean{inbibliography}{true}
\bibliographystyle{LHCb}
\bibliography{main}

\end{document}